\documentclass[twocolumn,pre,showpacs,floatfix,eqsecnum]{revtex4}

\usepackage{graphicx}%
\usepackage{dcolumn}
\usepackage{amsmath}

\newcommand{\bb}{\begin{equation}}
\newcommand{\en}{\end{equation}}
\newcommand{\xh}{\hat{x}}
\begin{document}

\title{State-dependent diffusion: thermodynamic consistency and its path integral formulation}

\author{A.W.C. Lau}
\affiliation{Department of Physics, Florida Atlantic University,
Boca Raton, Fl 33431}
\author{T.C. Lubensky}
\affiliation{Department of Physics and Astronomy, University of
Pennsylvania, Philadelphia, PA 19104}
\date{\today}

\begin{abstract}
The friction coefficient of a particle can depend on its position
as it does when the particle is near a wall.  We formulate the
dynamics of particles with such state-dependent friction
coefficients in terms of a general Langevin equation with
multiplicative noise, whose evaluation requires the introduction
of specific rules.  Two common conventions, the Ito and the
Stratonovich, provide alternative rules for evaluation of the
noise, but other conventions are possible. We show the requirement
that a particle's distribution function approach the Boltzmann
distribution at long times dictates that a drift term must be added to
the Langevin equation.  This drift term is proportional to the
derivative of the diffusion coefficient times a factor that
depends on the convention used to define the multiplicative noise.
We explore the consequences of this result in a number examples
with spatially varying diffusion coefficients.  We also derive
path integral representations for arbitrary interpretation of the
noise, and use it in a perturbative study of  correlations in a
simple system.

\end{abstract}
\pacs{05.40.-a} \maketitle

\section{Introduction}
\label{sec:intro}

Brownian motion provides a paradigm for exploring the dynamics of
nonequilibrium systems, especially those that are not driven too
far from equilibrium \cite{frey,vankampen1,risken,gardiner}. In
particular, the Langevin formulation of Brownian motion finds
applications that go beyond its original purpose of describing a
micron-sized particle diffusing in water. It has been extended to
treat problems in dynamics of critical phenomena \cite{justin}, in
glassy systems \cite{cugliandolo}, and even in evolutionary
biology \cite{biology}. Brownian motion is important for
soft-matter and biological systems because they are particularly
prone to thermal fluctuations \cite{frey,lubensky}, and Langevin
theory is an important tool for describing their properties, such
as the dynamics of molecular motors \cite{cell} and the
viscoelasticity of a polymer network \cite{morse}.

In most applications, the diffusion coefficient is assumed to be
independent of the state of the system. Yet, there are many
soft-matter systems in which the diffusion coefficient is state
dependent. A simple example of such a system is a particle in
suspension near a wall: its friction coefficient, and hence its
diffusion coefficient, depends because of hydrodynamic
interactions on its distance from the wall \cite{colloidfrench}, a
phenomenon that affects interpretation of certain single-molecule
force-extension measurements \cite{dna} and that plays a crucial role in
experimental verification of the fluctuation theorem in a dilute
colloidal suspension near a wall \cite{seifert}.  Similarly, the mutual
diffusion coefficient of two particles in suspension depends on
their separation \cite{quake}.  Other examples with
state-dependent diffusion include a particle diffusing in a
reversible chemical polymer gel \cite{bruinsma} and the dynamics
of fluid membranes \cite{cai}. In spite of the recent advances in
digital imaging methods to probe equilibrium properties of soft
matter \cite{crocker}, there have been relatively few experimental
studies of the dynamical properties of a physical system in which
the diffusion coefficient is state dependent. This is clearly an
area for further experimental exploration. Although the
mathematical problem of how to treat systems with state-dependent
diffusion has been studied for some time
\cite{vankampen1,morse,ermak,sancho,doi}, the results of these
studies have not been collected in one place to provide a clear
and concise guide to both theorists and experimentalists who might
use them.

In this largely expository paper, we develop a Langevin theory and
its associated path integral representation for systems with
state-dependent diffusion and explore its use in systems of
physical interest. In accord with previous treatments
\cite{vankampen1,risken,morse,colloidfrench,ermak,sancho,doi},
we show that a position-dependent diffusion
coefficient leads naturally to multiplicative noise.  This noise
is the product of a state-dependent prefactor proportional to the
square root of the diffusion coefficient and a state-independent
dependent Gaussian white noise function, and it is meaningless
without a prescription for the temporal order in which the two
terms are evaluated. There are two common prescriptions or
conventions for dealing with multiplicative noise: the Ito
convention in which the prefactor is evaluated before the Gaussian
noise and the Stratonovich convention which results when the
delta-correlated white noise is obtained as a limit of a noise
with a nonzero correlation time \cite{vankampen1,risken,gardiner}.
There are, however, other conventions as we will discuss.
Using general thermodynamic arguments, we show that in order for
Boltzmann equilibrium to be reached a drift term proportional to the derivative of the
diffusion coefficient times a factor depending on the convention
for the evaluation of multiplicative noise must be added to the
Langevin equation. Though this drift term has been noted before \cite{morse,ermak,sancho,doi},
we have found only one (recent) reference \cite{morse} that specifically
associate the form of the drift term with the convention for evaluating
multiplicative noise.  On the other hand, others claim that
it is the choice of the convention that is dictated by physics \cite{colloidfrench}.  In particular,
the authors of Ref.\ \cite{colloidfrench}, without allowing for the possibility of
the drift term, argued that neither Ito nor Stratonovich convention properly describes
the dynamics of a Brownian particle with a spatially varying friction coefficient, but
a third convention - what the authors called the isothermal convention, does.  Incidentally,
for this third convention, the drift term in our formalism vanishes.  Therefore,
the necessity of the drift term for enforcing thermal equilibrium is not widely known,
and it is often incorrectly ignored \cite{dna}.  Here, we aim to provide a clear exposition
for clarifying the technical issues that might have been a source of confusion in the literature.

This paper is organized as follows: in Sec.\ \ref{sec:formalism},
we first review the case of a uniform diffusion coefficient and
extend it to the case of spatially varying diffusion coefficient.
We discuss in depth the stochastic interpretation issues
associated with multiplicative noise, we derive the Fokker-Planck
equation, and we show that depending on the stochastic
interpretation, an additional drift term must be added to the
standard friction term in order for the system to relax to
equilibrium.  We also discuss how measurements of the eigenvalues
and eigenfunctions of the probability that a particle is at
position $x'$ at time $t + \delta t$ given that it was at position
$x$ at time $t$ can be used to obtain information about whether
the diffusion coefficient is state-dependent or not.  In Sec.\
\ref{sec:examples}, we present some exactly solvable toy models
that clearly illustrate the consequences of spatially varying
diffusion and suggest some experimental techniques which may
elucidate its role in colloidal tracking experiments. We also give
numerical confirmation that the extra drift term is needed to
produce equilibrium distribution. In Sec.\ \ref{sec:pathintegral},
we derive and discuss the path integral formulation for a Langevin
equation with a multiplicative noise, correlation functions, and
perturbation theory.  In Sec.\ \ref{sec:higher}, we briefly
summarize the results for multicomponent systems.  Technical
details are presented in the Appendices.

\section{Formalism in 1-d}
\label{sec:formalism}

\subsection{A review for the case of a uniform diffusion coefficient}
\label{sub:review}

Let us first briefly review the simplest case in which a Brownian
particle diffuses in space with a uniform diffusion constant \cite{vankampen1}.  In
the Langevin formulation of Brownian motion, the stochastic
equation of motion for the particle's position \cite{lubensky} is
\bb
\partial_t x = - \Gamma\,{ \partial {\cal H} \over \partial x } + g\,\eta(t),
\label{lang}
\en
where $x$ denotes the position, $\Gamma$ is the dissipative
coefficient (inverse mobility), ${\cal H}$ is the Hamiltonian, and
$g\, \eta(t)$ models the stochastic force arising from the rapid
collisions of the water molecules with the particle. The strength
of this force is set by $g$, and $\eta(t)$ is a Gaussian white
noise with zero mean, $\langle \eta(t) \rangle = 0$ and variance,
$\langle \eta ( t) \eta ( t') \rangle = \delta ( t -t' )$,
delta-correlated in time. The first term on the right hand side of
Eq.\ (\ref{lang}) describes a dissipative process. Thus, Eq.\
(\ref{lang}) can be viewed as a balancing equation in which the
first term drains the energy of the particle while the random
noise pumps it back. Equation (\ref{lang}) neglects an inertial
term that is only important at short times, typically less than
$10^{-7} \,\mbox{s}$ in soft-matter systems \cite{lubensky}. Thus,
Eq.\ (\ref{lang}) tacitly assumes that there is a separation of
time scales in which the time scale of the fast processes
reflecting microscopic degrees of freedom is much shorter than the
typical time scale for the random variable $x(t)$. Hence, the
white noise assumption in Eq.\ (\ref{lang}).

The Fokker-Planck equation \cite{vankampen1,risken,gardiner},
\bb
\partial_t P(x,t) = \partial_x \left [ \Gamma\,{ \partial {\cal H} \over \partial x }
+ {1 \over 2} \, g^2 \partial_x \right ] P(x,t),
\label{eq:FP}
\en
which can be derived for the Langevin equation, for the
probability density $P(x,t)$ that a particle is at position $x$ at
time $t$ provides an alternative to Langevin equation for
describing the motion of Brownian particles.  It is easy to see
that Eq.~(\ref{eq:FP}) has a steady state solution $P_s(x) \propto
\exp[ - {  2 \Gamma \over g^2 }\,{\cal H}]$. If a particle is in
equilibrium with a heat bath at temperature $T$, then $P_s(x)
\propto \exp[{  -\beta {\cal H}}]$ from which we conclude that
$g^2 = 2 \Gamma k_B T$.  If ${\cal H} = 0$, Eq.~(\ref{eq:FP})
reduces to a diffusion equation with diffusion constant $D =
g^2/2$.  Hence, for systems in equilibrium at temperature $T$, the
diffusion constant obeys the Einstein relation $D = k_B T \Gamma$.

\subsection{Extension to the case of state-dependent diffusion coefficient}
\label{sub:extension}

How must the Langevin equation for a Brownian particle be modified
when the friction coefficient $\Gamma$ depends on position $x(t)$,
{\em i.e.}, when $\Gamma$ depends on the state of the system.?
Though it is generally understood \cite{vankampen1,morse,ermak,sancho,doi,vankampen2}
that an $x$-dependent $\Gamma$ leads to an $x$-dependent $g$ and thus to
multiplicative noise $g\left [x(t)\right ] \eta(t)$, it is less well known that
the requirements of long-time thermal equilibrium require an
additional specific modification to the Langevin equation - the
addition of a convention-dependent drift term.  Though there are
discussions in the literature of this drift term \cite{morse,colloidfrench,ermak,doi},
they are not very detailed, and they generally treat only a
specific convention for dealing with multiplicative noise.  Here
we show that constraints of equilibrium require a unique drift
term with each noise convention and resolve any ambiguities
\cite{arnold} arising from the fact that multiplicative noise can
be interpreted in many ways.

Using the argument that the stochastic force is balanced by the
dissipative term as in the case of a uniform dissipative
coefficient above, we may reasonably postulate a Langevin
equation, which trivially generalizes Eq.\ (\ref{lang}) to the
case of spatially varying dissipative coefficient, to take the
following form:\bb
\partial_t x = - \Gamma(x)\,{ \partial {\cal H} \over \partial x } + g(x) \eta(t),
\label{langs}
\en where $g(x) = \sqrt{2 k_B T \Gamma(x)}$.  But we must first
confront the issue of interpreting the multiplicative noise $g(x)
\eta(t)$, which by itself is not defined
\cite{vankampen1,vankampen2}.  This is because the stochastic
nature of $\eta(t)$ which in general consists of a series of
delta-function spikes of random sign. The value of $g[x(t)]
\eta(t)$ depends on whether $g[x(t)]$ is to be evaluated before a
given spike, after it, or according to some other rule. It turns
out, as we will show shortly, that this naive generalization of
Eq.\ (\ref{lang}) to Eq.\ (\ref{langs}) is only valid for a
particular interpretation of the noise.

There are a number of approaches to assigning meaning to the
multiplicative noise, but they all boil down to providing rules
for the evaluation of the integral
\begin{eqnarray} {\cal J}(t,
\Delta t) =  \int_t^{t+\Delta t}ds\,g[\,x(s)]\, \eta(s),
\end{eqnarray}
in the limit of small $\Delta t$.  If $g(x)$ and $\eta(s)$ are
both continuous functions, this integral could, for arbitrary
$\Delta t$, be expressed via the first integral mean-value theorem as\bb
{\cal J}_{\mbox{\scriptsize cont}}(t, \Delta t) =  g[x(t_i)]
\int_t^{t+\Delta t}ds\,\, \eta(s)
\label{jcont}
\en where $t_i$ is a uniquely determined time in the interval $[t,
t+\Delta t]$.  In the limit of small $\Delta t$, this expression,
Eq.\ (\ref{jcont}), reduces trivially to $g[x(t)]\eta(t) \Delta
t$, to lowest order in $\Delta t$.  The noise $\eta(s)$ is,
however, not continuous and Eq.\ (\ref{jcont}) with a uniquely
determined time does not apply.  One can, however, use Eq.\
(\ref{jcont}) to motivate a definition of ${\cal J}(t, \Delta t)$
for a stochastic $\eta(s)$. There are two commonly used
conventions for defining ${\cal J}(t, \Delta t)$: the Stratonovich
convention
\bb {\cal J}_{S}(t) =  g\left [{ (x(t)+ x(t+\Delta
t))/2}\right  ]\,\int_t^{t+\Delta t}ds\, \eta(s), \en in which $g[
x(t) ]$ is evaluated at the midpoint of the interval $[
x(t),x(t+\Delta t)]$ and the Ito convention\bb {\cal J}_{I}(t) =
g[x(t)]\,\int_t^{t+\Delta t}ds\, \eta(s),
\label{ito}
\en
in which $g[ x(t) ]$ is evaluated before any noise in the
interval $(t, t +\Delta t)$ occurs. We will use a generalized
definition:\bb {\cal J}_{\alpha}(t,\Delta t) = g \left [\,\alpha
x(t+\Delta t) + (1 - \alpha) x(t)\,\right ] \,\int_t^{t+\Delta
t}ds\, \eta(s),
\label{alpha}
\en
which is parameterized by a continuous variable $\alpha \in
[0,1]$, that reduces to the Ito convention when $\alpha =0$, to
the Stratonovich convention when $\alpha = 1/2$, and to the isothermal convention
of Ref.\ \cite{colloidfrench} when $\alpha = 1$.

We note in passing that in the mathematics community, the Ito
calculus is most commonly used. Perhaps, this is because of the conceptual
simplicity arising from the property that the noise increment
$\int_t^{t+\Delta t} ds\,\eta(s)$ and $x(t)$ are statistically
independent as implied in Eq.\ (\ref{ito}), {\em i.e.} $\langle\,
g(x) \eta(t)\,\rangle = 0$ \cite{oksendal}. On the other hand, in
the physics community, the Stratonovich interpretation is favored.
In addition to the advantage that it gives rise to the ordinary
rules of calculus, the Stratonovich convention also has a deeper
physical origin. Since the noise term in Eq.\ (\ref{langs})
models, in a coarse-grained sense, the effects of microscopic
degrees of freedom that have finite (albeit short) correlation
times, this term should be physically interpreted as the limit in
which these correlation times go to zero. By the Wong-Zakai
theorem, this limit corresponds to a white noise that must be
interpreted using the Stratonovich convention \cite{oksendal}.
However, Eq.~(\ref{langs}) does not provide a correct description
for systems in constact with a thermal bath at temperature $T$ for
either interpretation: their associated Fokker-Planck equations do
not have long-time thermal-equilibrium solutions.

To return to our main discussion, it is clear that ${\cal
J}_{\alpha}(t,\Delta t)$ depends on the value of $\alpha$.
Integration of Eq.\ (\ref{langs}) yields
\bb \Delta x (t +\Delta
t) \equiv x(t+\Delta t ) - x(t) = {\cal J}_{\alpha}(t,\Delta t),
\en
when ${\cal H} =0$. The integral $\int_t^{t+\Delta t}ds\, \eta(s)$
is statistically of the order of $\sqrt{\Delta t}$, implying
$\Delta x (t +\Delta t)$ is also of the order of $\sqrt{\Delta
t}$.  Thus, $\alpha x(t+\Delta t) + (1 - \alpha) x(t) = x(t) +
\alpha \Delta x (t +\Delta t)$ has a term of order $\sqrt{\Delta
t}$ proportional to $\alpha$, and the order $\Delta t$ term in
${\cal J}_{\alpha}(t,\Delta t)$ depends on $\alpha$. An
alternative approach to defining ${\cal J}(t, \Delta t)$ is simply
to expand $g[x(s)]$ in the integrand as $g[x(s)] = g[x(t)]+ [x(s)
- x(t)]\,g'[x(t)]+ \cdots$. In this approach, which we outline in
Appendix \ref{app:alternative}, ambiguities in the interpretation
of ${\cal J}(t, \Delta t)$ are resolved by specifying the value of
the Heaviside unit step function, $\theta(t)$ at $t=0$.  Setting
$\theta(0) = \alpha$ is equivalent to using Eq.\ (\ref{alpha}) for
${\cal J}(t, \Delta t)$.

The stochastic integral ${\cal J}_{\alpha}(t, \Delta t)$ depends
on our convention for evaluating it, {\em i.e.} on $\alpha$.
Thus, different values of $\alpha$ define different dynamics. But
the requirements of thermal equilibrium should imply a unique
dynamics. What is missing? To resolve this dilemma, we consider
the general stochastic equation
\bb
\partial_t x = f(x) + g(x)\,\eta(t),
\label{spatial}
\en
where
\bb f(x) = - \Gamma(x)\,{ \partial {\cal H} \over
\partial x } + f_1(x),
\label{eq:f_1}
\en
in which we leave $f_1(x)$ unspecified
for the moment. Eq.\ (\ref{spatial}) is easily integrated using
the rules we just outlined to yield
\begin{eqnarray}
x(t + \Delta t ) - x(t) &=& \int_t^{t + \Delta t}ds \,
\left \{ f[x(s)] + g[x(s)]\, \eta(s) \right \}\nonumber \\
&=& f[ x(t) + \alpha \Delta x] \Delta t  \nonumber \\
&+& g[ x(t) + \alpha \Delta x] \,\int_t^{t+\Delta t}ds\, \eta(s),
\label{x}
\end{eqnarray}
from which we obtain, to the first order in $\Delta t$,
\begin{eqnarray}
\langle \Delta x \rangle &=& f(x_0)\,\Delta t
+ \alpha g(x_0)g'(x_0) \Delta t, \\
\langle ( \Delta x )^2 \rangle &=& g^2(x_0) \Delta t,
\label{diffconst}
\end{eqnarray}
where we set $x(t) =x_0$.  Thus, there is a stochastic
contribution, $\alpha g g' \Delta t$, to $\langle \Delta x
\rangle$ arising from the $x$ dependence of $g$ and depending on
the convention for evaluating ${\cal J}(t, \Delta t)$. In
equilibrium, $\langle \Delta x \rangle$ should be independent of
$\alpha$. Thus, it is apparently necessary to include a
contribution to $f(x)$ depending on $\alpha$.

\subsection{Derivation of the Fokker-Planck Equation and Equilibrium Conditions}
\label{subsec:fpe}

To determine the appropriate form of $f(x)$ and $g(x)$ to describe
equilibrium systems with a spatially varying friction coefficient
$\Gamma(x)$, we derive the Fokker-Planck equation for the
probability density $P(x,\,t )$. The Fokker-Planck equation is most
easily derived using the identity\bb P(x ,t +\Delta t) = \int
dx_{0}\,P(x ,t + \Delta t | x_{0}\,t )P(x_{0}\,t ),
\label{ck}
\en where $P(x ,t + \Delta t | x_{0}\,t )$ is the conditional
probability distribution of $x$ at time $t + \Delta t$ given that
it was $x_{0}$ at time $t$.  It is defined by\bb P(x ,t + \Delta t
| x_{0}\,t ) = \left \langle \delta[x - x(t + \Delta t)]  \right
\rangle_{x_0,t} \en where the average is over the random noise
$\eta(s)$ and $x(t+\Delta t)$ is determined by Eq.\ (\ref{x}) with
$x(t) = x_0$.  Taylor expanding the conditional probability around
$x_0$ yields \begin{eqnarray}
P(x ,t + \Delta t | x_{0}\,t ) &=&
\delta(x - x_0) - \langle \Delta x \rangle \, { \partial  \over \partial x}\,\delta(x - x_0) \nonumber \\
&+& {1 \over 2 }\,\langle ( \Delta x )^2 \rangle \,{ \partial^2
\over \partial x^2}\,\delta(x - x_0) +\cdots. \nonumber
\end{eqnarray}
\begin{widetext}
Then using this in Eq.\ (\ref{ck}), we obtain\begin{eqnarray}
\partial_t P(x,t) &=& {\partial \over \partial x }  [- f(x) -\alpha g(x) g'(x) ] P(x,t)
+  {1 \over 2} {\partial^2 \over \partial x^2 } \left [\,g^2(x) P(x,t) \right ] \label{fpe1} \\
&=& {\partial \over \partial x } \left [ \Gamma(x) {\partial
{\cal H} \over \partial x} - f_1(x) +(1-\alpha ) g(x)g'(x) + {
1\over 2}\, g^2(x) {\partial \over \partial x } \right ] P(x,t).
\end{eqnarray}
\end{widetext}
For an equilibrium system, this equation must have a steady state
solution with the canonical form \bb P(x,t) \sim e^{- {\cal
H}/(k_B T)} \en that is always approached at long times. Such a
solution is guaranteed if\begin{eqnarray}
g^2(x) &=& 2 k_B T \Gamma(x), \\
f_1(x) &=& (1- \alpha ) g(x)g'(x) = 2 (1- \alpha ) k_B T \Gamma'(x).\,\,\,\,\,\,\,\,\,\,
\label{fone}
\end{eqnarray}
Thus, an additional drift term, $f_1(x)$, which depends on the
convention for evaluating ${\cal J}(t, \Delta t)$, must be added
to the standard friction term, $- \Gamma(x) \partial_x {\cal H}$,
in the equation for $\partial_t x$ in order for the system to
evolve to the Boltzmann distribution at long times, {\em i.e.}, be
consistent with thermodynamics. Note that $f_1(x)$ is proportional
to the temperature $T$, indicating that its origin arises from
random fluctuations rather from forces identified with a
potential. It is clear now from Eq.\ (\ref{fone}) that if we
insist on using the Langevin equation in the form of Eq.\
(\ref{langs}), we are forced to take $\alpha =1$
\cite{colloidfrench}.

It is customary to express the Fokker-Planck equation in terms of
the diffusion constant rather then the friction coefficient. From
Eq.\ (\ref{diffconst}) for $\langle ( \Delta x )^2 \rangle $, we
can identify $g^2(x)$ with the short-time diffusion constant $D(x)
= 2 k_B T \Gamma(x)$. With this definition of $D(x)$ and $f_1(x)$
given by Eq.\ (\ref{fone}), the Fokker-Planck equation becomes\bb
\partial_t P(x,t) = {\partial \over \partial x }\,D(x) \left [
\beta {\partial {\cal H} \over \partial x}  + {\partial \over
\partial x }  \right ] P(x,t),
\label{fpthermal}
\en where $\beta = 1/(k_B T)$.  As required, this equation is
independent of $\alpha$: different conventions now give the same
equilibrium condition as they should. For a free particle
diffusing in spatially varying $D(x)$, ${\cal H} = 0$ and Eq.\
(\ref{fpthermal}) becomes \bb
\partial_t P(x,t) = {\partial \over \partial x }\,D(x) {\partial \over \partial x } P(x,t).
\en This implies that the correct generalization of Fick's Law for
equilibrium systems with a spatially-varying diffusion coefficient
is given by\bb J(x,t) = - D(x) \partial_x P(x,t).
\label{fix}
\en Historically, the generalization of Fick's law has long been
debated \cite{mark}. It is commonly acknowledged that Eq.\
(\ref{fix}) is right even though many derivations to the right of
side of Eq.\ (\ref{fix}) seem not to be as transparent as the one
given above.

\subsection{Experimental probes of $D(x)$}

One interesting property of Eq.\ (\ref{fpthermal}) is that it
necessarily has an eigenstate with eigenvalue zero and
eigenfunction given by the equilibrium distribution the $P_{eq}(x)
\propto e^{- \beta {\cal H}(x)}$.  This fact is exploited by
Crocker et al.\ \cite{crocker} to measure directly the interaction
between an isolated pair of colloidal particles.  In these
experiments, the data from tracking the motion of the particles
are used to compute the conditional probability $P(x, t
+\delta t | x' t)$, which may be viewed as the Green's function to
or the inverse of the Fokker-Planck equation, Eq.\
(\ref{fpthermal}).  The equilibrium distribution is then the
solution to\bb
P_{eq}(x) = \int dr'\,P(x,\,t +\delta t\,|\,x',\,t) P_{eq}(x'),
\en
from which the interaction
potential can be constructed via $U(x) = - k_B T \log P_{eq}(x)$.

Since the conditional probability contains all the dynamical
information of the system, one could in principle characterize how
the system relaxes to equilibrium by extracting the nonzero
eigenvalues of Eq.\ (\ref{fpthermal}).  In particular, the
Fokker-Planck equation describing a system with a state-dependent
diffusion coefficient would have eigenvalues and eigenfunctions
that are, in general, different from those of a system with a
uniform diffusion coefficient, even though the two systems have
the same Hamiltonian. Thus, in principle, one could extract the
Hamiltonian from image analysis following the procedures of
Crocker \cite{crocker}, and solve Eq.\ (\ref{fpthermal}) with
uniform diffusion constant to obtain a set of eigenvalues
(probably numerically) and compare it with experimentally measured
eigenvalues, which can be extracted from the measured conditional
probability.  If they are different, then the diffusion
coefficient is state dependent, and one needs to model the
diffusion coefficient to understand the dynamical behaviors of the
system.  We suggest this procedure as a possible general method for
experimentalists to explore the dynamics and measure the state
dependent friction coefficient in, for example, hydrodynamic
interactions between two spheres \cite{crocker2}, diffusion of
particles in polymer solution \cite{verma}, and rods in a nematic
environment \cite{dogic}.

\section{Illustrative Examples}
\label{sec:examples}

In this section, we consider some exactly solvable toy models to
illustrate some central ideas presented in the last section.  In
particular, we address the effects of spatial dependence in the
diffusion coefficient and use numerical solution of the Langevin
equation to show that equilibrium distribution is obtained only if
$f_1(x)$, Eq.\ (\ref{fone}), is added to the standard friction
term.

\subsection{Diffusion of a particle near a wall}
\label{subsec:particlenearwall}

How does a diffusion coefficient acquire a spatial dependence? The
simplest example is a Brownian particle diffusing near a wall
located at $z=0$. Brenner \cite{brenner} has shown that for $z>0$
the diffusion coefficient acquires a spatial dependence in which
it is zero at the wall, rises linearly in $z$, and approaches a
uniform bulk value of $D_0$ at large $z$ as
\bb { D(z) \over
D_0 } = 1- { 9 \over 8} {a \over z} + \ldots \,\, .
\label{wall}
\en
Note the long-range component of $D(z)$ in Eq.\ (\ref{wall}),
which reflects the long-ranged nature of the hydrodynamic
interaction.  Recently, it has been pointed out that in single
molecule experiments, it is crucial to take the spatial dependence
in the diffusion coefficient properly into account \cite{dna}.

Rather than to treat the system with the above $D(z)$, we
consider a toy model, which correctly describes diffusion close to
a wall, in which $D(z) = \Lambda z $. This diffusion coefficient
has another experimental realization: diffusion of a colloidal
particle bounded by two parallel walls, with one of the walls
slightly tilted \cite{colloidfrench}. Then, the diffusion
coefficient acquires a spatial dependence, approximately given by
$D(z) \sim z $, for the motion of the particle parallel to the
walls. In this case, the Fokker-Planck equation becomes
\bb
\partial_t P(z,t) =\Lambda {\partial \over \partial z }
\,z\,{\partial \over \partial z }\,P(z,t),
\label{linear}
\en
which can be solved exactly.  Let $ P(z,t) = \sum_n c_n e^{-
\lambda_n t} \psi_n(z),$ where $\lambda_n$ are a set of the
eigenvalues.  With the transformation $y = \sqrt{z}$, Eq.\
(\ref{linear}) can be written as\bb \psi_n^{''}(y) + {1 \over y}
\psi_n^{'}(y) + { 4 \lambda_n \over \Lambda} \psi_n(y) = 0, \en
whose solution is the Bessel function: $\psi(y) = J_0( k y )$, and
whose eigenvalues form a continuous spectrum given by $\lambda =
\Lambda k^2/4$.  The probability distribution as a function of
time can be written as
\bb
P(z,t) = \int_0^{\infty} dk\,c(k)\,e^{-
\Lambda k^2 t/4}\,J_0( k \sqrt{z} ).
\en
The probability distribution for a particle at $z=z_0$ at time
$t=0$ evolves as
\bb
P(z,t) = {1
\over \Lambda t}\,e^{- \left [{ (z +z_0) /( \Lambda t)} \right ] }
I_0\left[\,2\sqrt{z z_0}/(\Lambda t) \right ].
\en
Unlike its counterpart for a uniform diffusion coefficient, this
probability distribution is non-Gaussian. It is straightforward to
calculate the moments:
\begin{eqnarray}
\langle z(t) \rangle &=& z_0 + \Lambda t \nonumber \\
\langle \left [ z(t) -z_0 \right ]^2 \rangle &=& 2 \Lambda t\,(
z_0 + \Lambda t). \nonumber
\end{eqnarray}
These behaviors are very different from those of a constant
diffusion. In particular, the mean-squared displacement exhibits
ballistic behavior.  It is interesting to observe that the second
moment can be written as $\langle (\, z(t) - z_0)^2\, \rangle =
2\,\Lambda \langle z(t) \rangle\,t$. This suggests that in order
to extract the diffusion coefficient for this simple problem, we
need to know not only the second moment $\langle (\, z(t) - z_0)^2
\rangle $, but also the first moment $\langle z(t) \rangle$.  Only
for short times does the second moment reduce to $\langle (\, z(t)
- z_0)^2\, \rangle \sim 2\,\Lambda z_0\,t = 2 D(z_0) t$, which is
the formula commonly used to extract the diffusion coefficient. It
is clearly incorrect to use this formula for times greater than
$z_0/\Lambda$.  The method we suggested at the end of the last
section compliment this approach. Note also that the $\langle z(t)
\rangle \sim t $ behavior has been measured in Ref.\
\cite{colloidfrench}.

\begin{figure}
    \resizebox{2.7in}{!}{\includegraphics{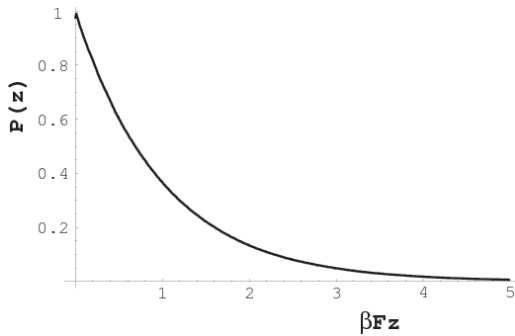}}
    \caption{\label{fig:disforce} Stationary distribution for a particle diffusing in a diffusion
    coefficient $D(z)= \Lambda z$ subject to a constant force ${\bf F} = - F\,\hat{z}$.  What is
    shown here is the numerical simulation of the Langevin equation Eq.\ (\ref{force});
    it is of the form $e^{-\beta F z}$, as expected.}
\end{figure}

If the particle is subject to constant force $F$, like gravity, in
the $-z$ direction, then the Fokker-Planck equation is\bb
\partial_t P(z,t) = \Lambda {\partial \over \partial z }\,z \left [
\beta F  + {\partial \over \partial z }  \right ] P(z,t). \en This
problem can also be solved exactly.  Let $ P(z,t) = e^{-\beta F
z}\,\sum_n c_n e^{-\lambda_n t}\psi_n(z),$ we find that the
eigenfunctions satisfy the Laguerre equation
\bb x\psi^{''} + (1-x)
\psi^{'} + { \lambda_n \over \Lambda \beta F}\,\psi =0,
\en
with eigenvalues $\lambda_n = n \Lambda \beta F$. The eigenvalue
spectrum is discrete rather than continuous as it is in the case
of a constant diffusion coefficient. If the particle is initially
at $z_0$, the distribution evolves as
\begin{eqnarray}
P(z,t) &=& {
\beta F \over 1 - e^{-\Lambda \beta F t}} \exp{- \left [ { \beta F
( z + z_0 e^{-\Lambda \beta F t}) \over 1 - e^{-\Lambda \beta F
t}} \right ]}
\, \nonumber \\
&\times& I_0 \left [ { 2 \beta F \sqrt{ z_0 z e^{-\Lambda \beta F
t }} \over 1 - e^{- \Lambda \beta F t}} \right ].
\end{eqnarray}
Note that at $t \rightarrow \infty$, this distribution reaches the
equilibrium distribution $P_{eq} \sim e^{-\beta F z }$. The first
two moments of $z(t)$ are
\begin{eqnarray} \langle
z(t) \rangle &=& z_0 e^{-\Lambda \beta F t} + { 1\over \beta F}
\left ( 1 - e^{- \Lambda \beta F t} \right ) \nonumber \\
\langle z^2(t) \rangle &=& z_0^2 e^{- 2 \Lambda \beta F t} +
{ 4 z_0 e^{-\Lambda \beta F t} \over \beta F} ( 1 - e^{-\Lambda \beta F t} ) \nonumber \\
&+& { 2 \over (\beta F)^2 } ( 1 - e^{-\Lambda \beta F t} )^2.
\end{eqnarray}
Note that at long time $\langle z(t) \rangle = k_BT/F$ as thermal
equilibrium dictates.

We numerically solve the Langevin equation corresponding this
problem\bb
\partial_t z(t) = \Lambda + \Lambda \beta F z + \sqrt{2 \Lambda z}\,\eta(t),
\label{force}
\en in the Ito convention \cite{numerics}.  Note that the first
term in the right-hand side arises from the additional drift,
$f_1(x)$ given by Eq.\ (\ref{fone}).  The result for the
stationary distribution is plotted in Fig.\ \ref{fig:disforce}.
Obviously, it agrees with the equilibrium distribution $P_{eq}
\sim e^{-\beta F z }$.

\subsection{Diffusion of a particle bounded by two parallel walls}
\label{particletwowalls}

\begin{figure}
    \resizebox{2.5in}{!}{\includegraphics{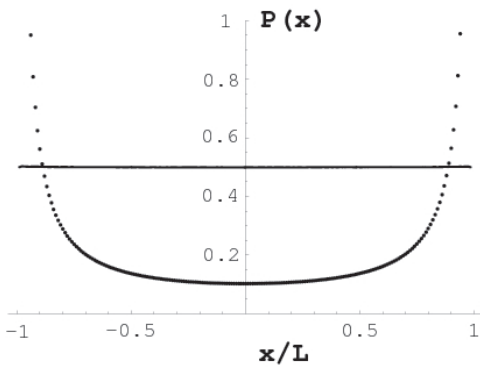}}
    \caption{\label{fig:twowalls} Stationary distribution for a particle diffusing between
    two walls with a diffusion coefficient $D(x) = D_0 \left [ 1 - (x/L)^2 \right ]$.  The
    solid line is the numerical simulation of the Langevin equation in Eq.\ (\ref{twowalls})
    and the dotted line is the numerical simulation of the Langevin equation without the extra
    $gg'$ term.}
\end{figure}

Next, we consider the diffusion of a particle bounded by two
walls, which was studied experimentally in Ref.\ \cite{libchaber}
and more recently in Ref.\ \cite{grier}.  We approximate the
spatially varying diffusion coefficient of this system by $D(x) =
D_0 \left [ 1 - (x/L)^2 \right ]$. The resulting  Fokker-Planck
equation is
\bb
\partial_t P(x,t) = D_0 {\partial \over \partial x }\,\left [ 1 - (x/L)^2 \right ]
\,{\partial \over \partial x }\,P(x,t),
\en
with boundary conditions that particles cannot penetrate the
walls, {\em i.e.}, that the flux at both walls be zero:
\bb
J(x,t) = D_0 \left [ 1 - (x/L)^2 \right ] \,{\partial \over
\partial x }\,P(x,t) =0 \,\,\text{at} \,\, x = \pm L .
\en
Again the solution to this problem differs considerably from that
with a spatially uniform diffusion coefficient.  The spectrum is
discrete rather than continuous with eigenvalues $\lambda_n =
n(n+1)D/L^2$ ($n= 1, 2, ...$) and the associated eigenfunctions
are Legendre polynomials rather than linear combinations of plane
waves. The first two moments of $x(t)$ are
\begin{eqnarray}
\langle x(t) \rangle &=&  x_0 e^{-2D_0t/L^2} \nonumber \\
\langle x(t)^2 \rangle &=& x_0^2 e^{- 6 D_0 t/L^2} + { L^2 \over
3} \left ( 1 - e^{-6D_0t/L^2} \right ).
\end{eqnarray}
These moments again are different from the case in which the
diffusion is uniform.

We performed numerical simulation of the Langevin equation
\bb
\partial_t x = - 2D_0 x /L^2 + \sqrt{2 D_0 ( 1- x^2/L^2)}\,\eta(t),
\label{twowalls}
\en
where the first term arises from the $gg'$ term. In Fig.\
\ref{fig:twowalls}, we plot the long time distribution (solid
line) which is uniform as it should be.  We also show the
numerical results for the case in which we did not add the $gg'$
(dotted line). Clearly, we get the wrong answer if we do not add
the $gg'$ term.

\subsection{Diffusion constant: $D(x) = D_0 ( 1+  \gamma \,x^2 )$ }
\label{subsec:wierd}

As a final example, let us consider a free particle diffusing with
$D(x) = D_0 ( 1 + \gamma\, x^2 )$ in the bulk. The Fokker-Planck
equation is given by
\bb
\partial_t P(x,t) = D_0 {\partial \over \partial x }\,\left [ 1 + \gamma\,x^2 \right ]
\,{\partial \over \partial x }\,P(x,t).
\en
Multiplying both sides by $x^2$ and integrating, we find
\bb
\partial_t \langle\,x^2(t) \rangle = 2 D_0 + 6 D_0 \gamma\,\langle\,x^2(t) \rangle ,
\en
whose solution is
\bb
\langle\,x^2(t) \rangle = { 1\over 3\,
\gamma } \left [\,\exp ( 6 D_0 \gamma t ) -1\, \right ].
\en
Thus, the second moment grows exponentially with time; this
peculiar behavior illustrates the dramatic effects of the noise in
problems with a spatial dependent diffusion coefficient.

\section{Path-integral Formulation}
\label{sec:pathintegral}

Path-integral formalisms provide an alternative to the
Fokker-Planck and Langevin equations for the description of
stochastic dynamics. They have the advantage that well-established
perturbative and non-perturbative field-theoretic techniques
\cite{justin,MSR} can be used to calculate the effects of
nonlinearities. They also provide a convenient treatment of
correlation and response functions. The path integral for a
state-dependent dissipation coefficient has been derived
previously either in the Stratonovich or Ito convention
\cite{justin,arnold,phythian,graham}.  In this section, we derive
the path-integral for the general $\alpha$ convention, use it
along with the detailed balance, a condition that any thermal
systems must satisfy, to shed further insight into the additional
drift term derived in Sec.\ \ref{subsec:fpe}. We also discuss
equilibrium correlation and response functions and prove the
Fluctuation-dissipative theorem for state-dependent diffusion
coefficient. We then set up perturbation theory for systems with a
coordinate-dependent friction coefficient.

The path-integral is based on the statistics of a path $x(t)$. We
discretize the path into segment $x_i = x(t_i)$ with $t_0 < t_1 <
\cdots < t_{N-1} < t_f$ and $\Delta t \equiv t_n - t_{n-1}$ small.
The joint probability distribution, $P(x_N\,t_N; x_{N-1}\,t_{N-1};
\ldots ;x_1\,t_1| x_0\,t_0)$ that $x(t)$ takes on values of $x_1$
at time $t_1$, $x_2$ at time $t_2$ and so on, given that it has
value of $x_0$ at time $t_0$, is then
\begin{eqnarray}
&&P(x_f\,t_f; x_{N-1}\,t_{N-1}; \ldots ;x_1\,t_1| x_0\,t_0)  \nonumber \\
&& = \langle \, \delta[x_N - \phi(t_N;x_0,t_0)]\ldots \delta[x_{1}
- \phi(t_{1};x_0,t_0)]\,\rangle, \nonumber
\end{eqnarray}
where the average is taken with respect to the noise and
$\phi(t_i;x_{i-1},t_{i-1})$ is the solution to the Langevin
equation, Eq.\ (\ref{spatial}), for $x(t_i)$ given that
$x(t_{i-1}) = x_{i-1}$. Since the noise in Eq.\ (\ref{spatial}) is
delta-correlated in time, the noise in different time intervals is
not correlated, and $x_i$ depends only on $x_{i-1}$. Thus, we can
write
\begin{eqnarray} P(x_N\,t_N; \ldots ;x_1\,t_1|
x_0\,t_0) = \prod_{i=1}^{N} \langle \delta[x_i -
\phi(t_i;x_{i-1},t_{i-1})]\rangle. \nonumber
\end{eqnarray}
The function \bb P(x_i\,t_i | x_{i-1}\,t_{i-1} ) = \langle
\delta[x_i - \phi(t_i;x_{i-1},t_{i-1})]\rangle
\label{markov}
\en
gives the conditional probability that the random variable
$x(t)$ has the value $x_i$ at time $t_{i}$ given that it had a
value $x_{i-1}$ at $t_{i-1}$. Using Eq.\ (\ref{markov}) and the
identity $P(x\,t) = \int\,dx'\,P(x\,t;x'\,t'),$ it is easy to see
that
\begin{eqnarray}
&& P(x_i\,t_i | x_{i-2}\,t_{i-2} ) = \int dx_{i-1}\,P(x_i\,t_i | x_{i-1}\,t_{i-1} )  \nonumber \\
&& \hphantom{pppppppppp}\times P(x_{i-1}\,t_{i-1} | x_{i-2}\,t_{i-2} ), \label{ck1} \\
&& P(x_i\,t_i) = \int dx_{i-1}\,P(x_i\,t_i | x_{i-1}\,t_{i-1}
)P(x_{i-1}\,t_{i-1} ).
\label{ck2}
\end{eqnarray}
Equation (\ref{ck1}) is the Chapman-Kolmogorov equation, which
defines a Markov process \cite{vankampen1}, while Eq.\ (\ref{ck2})
is just an identity, true for all stochastic processes.  Note that
a Markov process is completely specified if we know $P(x_i\,t_i)$
and $P(x_i\,t_i | x_{i-1}\,t_{i-1} )$, but they are not arbitrary
because they are linked through Eqs.\ (\ref{ck1}) and (\ref{ck2}).
Using Eq.\ (\ref{ck1}), the conditional probability for the
particle to go from $x_0$ at time $t_0$ to $x_f$ at time $t_f$
is
\begin{eqnarray}
&& P(x_f\,t_f| x_0\,t_0) = \int dx_{N-1}\ldots\int dx_{1} \, \times \,\label{products} \\
&& \langle \delta[x_f - \phi(t_f;x_{N-1},t_{N-1})]\rangle \ldots
\langle \delta[x_{1} - \phi(t_{1};x_0,t_0)]\rangle. \nonumber
\end{eqnarray}
This is the basic construct for the path integral.  First, we have
to evaluate $\langle \delta[x_i -
\phi(t_i;x_{i-1},t_{i-1})]\rangle$. We discretize Eq.\
(\ref{spatial}) as follows:
\bb x_i = x_{i-1} + \Delta t \,f_i +
g_i \int_{t_{i-1}}^{t_i} dt' \, \eta(t'),
\label{dis}
\en
where $\Delta t = t_i - t_{i-1}$, $f_i = f[\,\alpha\,x_i +
(1-\alpha)\,x_{i-1}] $, and $g_i = g[\,\alpha\, x_i +
(1-\alpha)\,x_{i-1}] $. We next introduce the function $h(x_i,
x_{i-1})$:\bb h(x_i, x_{i-1}) = { x_i - x_{i-1} - \Delta t\, f_i
\over g_i } - \int_{t_{i-1}}^{t_i} dt' \, \eta(t'), \en which
vanishes when $x_i$ is the unique solution to the Eq.\ (\ref{dis}),
$\phi(t_i;x_{i-1},t_{i-1})$, {\em i.e.},
$h[\phi(t_i;x_{i-1},t_{i-1}), x_{i-1}] = 0$. Using the property of
the delta function \bb \delta[h(x_i, x_{i-1})] =  \left | {
\partial h \over  \partial x_i } \right |_{x_i = \phi(t_i)}^{-1}
\,\delta[x_i - \phi(t_i)], \nonumber \en and noting that $\left |
{ \partial h \over  \partial x_i } \right |_{x_i =
\phi(t_i)}^{-1}$ depends only on $x_i$ and $x_{i-1}$, which are
set by the delta function and not explicitly on the noise, we have
\begin{eqnarray} \langle \delta[h(x_i, x_{i-1})]\rangle  &=& \left
\langle  \left | { \partial h \over  \partial x_i }
\right |_{x_i = \phi(t_i)}^{-1} \,\delta[x_i - \phi(t_i)] \right \rangle \nonumber  \\
&=& \left | { \partial h \over  \partial x_i } \right |^{-1}\,
\langle \delta[x_i - \phi(t_i)] \rangle,\nonumber
\end{eqnarray}
since for any function $q[\phi(t_i)]$, $ \langle q[\phi(t_i)]
\delta[x_i -\phi(t_i) ] \rangle = q(x_i) \langle \delta[x_i
-\phi(t_i) ] \rangle $.  We can, therefore, write the conditional
probability as\begin{eqnarray}
P(x_i t_i |x_{i-1} t_{i-1}) &=& \langle \delta[x_i - \phi(t_i;x_{i-1},t_{i-1})] \rangle \nonumber \\
& =& \left | { \partial h(x_i, x_{i-1}) \over  \partial x_i }
\right | \, \langle \delta[h(x_i, x_{i-1})] \rangle, \nonumber
\end{eqnarray}
with\begin{eqnarray} { \partial h \over  \partial x_i } = {1 \over
g_i} \left [ { 1 - \alpha \Delta t \,f_i' - \alpha\, { g_i' \over
g_i }\left ( x_i - x_{i-1} - \Delta t \, f_i \, \right ) } \right
], \nonumber
\end{eqnarray}
where $'$ denotes the derivative.  The average over noise can be
easily done with the aid of the Fourier representation of the
delta function: \begin{eqnarray} \langle \delta[h(x_i, x_{i-1})]
\rangle  &=& \int {d k_i \over 2 \pi } \,
e^{+ \imath { k_i  \over g_i }( x_i - x_{i-1} - \Delta t f_i )}  \nonumber \\
&\times& \left \langle e^{ - \imath k_i \int_{t_{i-1}}^{t_i} dt'
\, \eta(t')} \right \rangle,
\nonumber \\
&=&\int {d k_i \over 2 \pi }\, e^{+ \imath { k_i  \over g_i }( x_i
- x_{i-1} - \Delta t f_i )  - { 1 \over 2} k_i^2 \Delta t},
\nonumber
\end{eqnarray}
where we have made use of the fact that $\int_{t_{i-1}}^{t_i} dt'
\, \eta(t')$ is a zero-mean Gaussian random variable with variance
$\Delta t$. Putting these results together, we can express $P(x_i
t_i |x_{i-1} t_{i-1})$ as
\begin{eqnarray}
&& P(x_i t_i |x_{i-1} t_{i-1}) = \int {d k_i \over 2 \pi g_i }\,
e^{+ \imath { k_i  \over g_i }( x_i - x_{i-1} - \Delta t f_i )  - { 1 \over 2} k_i^2 \Delta t} \nonumber \\
&& \phantom{pppp}\times \left [ 1 - \alpha \Delta t \,f_i' -
\alpha\, { g_i' \over g_i }\left ( x_i - x_{i-1} - \Delta t \, f_i
\, \right ) \right ].\,
\label{path1}
\end{eqnarray}

Next, in order to derive the path integral which is of the form
$\sim e^{- {\cal S}}$, we need to ``exponentiate" the bracket term
in Eq.\ (\ref{path1}) and keep all the terms that are of order of
$\Delta t$ in the exponential.  However, we cannot simply
exponentiate the third term in the bracket because this term
contains $\Delta x_i \equiv x_i-x_{i-1}$, which is of order of
$\sqrt{\Delta t}$.  This is noted in Ref.\ \cite{arnold}, where
the author derives the path integral for the Stratonovich
convention, and circumvents this difficulty by keeping the second
order term in $\Delta x$  in the exponential and replacing this
term with its average value. Although the final expression is
correct, that derivation might be inconsistent with the concept of
path integral since that derivation is valid only in the
mean-squared sense instead of for all paths, as required by the
path integral. Here, we provide an alternative derivation that is
valid for each path.  First, we note that the last term in the
bracket can be written as
\begin{widetext}
\begin{eqnarray}
- \int {d k_i \over 2 \pi g_i }\,\left [\,\alpha { g_i' \over g_i
} \left ( \Delta x_i - \Delta t \, f_i \, \right )\right ]\, e^{+
\imath { k_i  \over g_i }( \Delta x_i - \Delta t f_i )  - { 1
\over 2} k_i^2 \Delta t} &= & -\,\alpha g_i' \int {d k_i \over 2
\pi g_i }\,e^{- { 1 \over 2} k_i^2 \Delta t} \left (-\imath
{\partial \over \partial k_i} \right )
e^{+ \imath { k_i  \over g_i }( \Delta x_i - \Delta t f_i )} \nonumber \\
&=& +\,\alpha g_i'  \int {d k_i \over 2 \pi g_i } \left [\, \imath
k_i \Delta t\,\right ] e^{+ \imath { k_i  \over g_i }( \Delta x_i
- \Delta t f_i )  - { 1 \over 2} k_i^2 \Delta t},
\end{eqnarray}
where the last line explicitly of order of $\Delta t$ and can,
therefore, be exponentiated without incurring any error to the
first order in $\Delta t$. Returning to the conditional
probability, we have
\begin{eqnarray} P(x_i t_i |x_{i-1} t_{i-1})
&=& \int {d k_i \over 2 \pi g_i }\, e^{+ \imath { k_i  \over g_i
}( \Delta x_i  - \Delta t f_i )  - { 1 \over 2} k_i^2 \Delta t}
\left [\,1 - \alpha \Delta t \,f_i' + \imath \alpha \Delta t k_i g_i' \,\right ] \\
&=&\int {d k_i \over 2 \pi g_i }\, e^{+ \imath { k_i  \over g_i }(
\Delta x_i  - \Delta t f_i  + \alpha \Delta t g_i g_i')  - { 1
\over 2} k_i^2 \Delta t
- \alpha \Delta t \,f_i' } \label{conditional2}\\
&=& { 1 \over \sqrt{ 2 \pi \Delta t } \, g_i }\, e^{- { \Delta t
\over 2 g_i^2 } \left [ { \Delta x_i \over \Delta t } - f_i +
\alpha g_i g_i' \right ]^2 - \alpha \Delta t f_i'},
\label{conditional}
\end{eqnarray}
where the last line is valid to the first order in $\Delta t$. It
should be noted that the Fokker-Planck equation, Eq.\
(\ref{fpe1}), can also be derived using Eq.\ (\ref{conditional})
and the identity of Eq.\ (\ref{ck2}).  This is done in the Sec.\
\ref{sub:derivation}. Returning to Eq.\ (\ref{products}), we have
\begin{eqnarray}
P(x_f\,t_f| x_0\,t_0) &=& \int { dx_{1} \over \sqrt{ 2 \pi \Delta
t } g_1 }\ldots \int { dx_{N-1} \over \sqrt{ 2 \pi \Delta t }
g_{N-1} } { 1 \over \sqrt{ 2 \pi \Delta t } g_N}  e^{- \sum_i
{\Delta t \over 2 g_i^2 } \left
[ { x_i - x_{i-1} \over \Delta t } - f_i + \alpha g_i' g_i  \right ]^2 - \sum_i \alpha \Delta t f_i'} \\
&=& \int_{x_0}^{x_f}{\cal D}x \,e^{-\int_{t_0}^{t_f}dt \,\left \{
{1 \over 2 g(x)^2 } \left [ \partial_t x - f(x) +
\alpha\,g(x)\,g'(x) \right ]^2 + \alpha f'(x) \right \}} =
\int_{x_0}^{x_f}{\cal D}x \,e^{-{\cal S}},
\end{eqnarray}
with ${\cal D}x \equiv \prod_{i=1}^{N} { dx_{i} \over \sqrt{ 2 \pi
\Delta t } g_i }$, and the action given by
\bb {\cal S} =
\int_{t_0}^{t_f}dt \left \{ {1 \over 2 g(x)^2 } \left [ \partial_t
x - f(x) + \alpha\,g(x)\,g'(x) \right ]^2 + \alpha f'(x)\right \},
\label{pathspatial}
\en
where we have taken the formal limit of by letting $N \rightarrow
\infty$ and $\Delta t \rightarrow 0$. Note the extra terms in the
${\cal S}$ coming from the Jacobian $\left |\partial h/\partial x_i \right |$;
they are needed in order to ensure that $\int dx_f
P(x_f\,t_f|x_0\,t_0) =1$. This can be demonstrated by explicit,
but tedious, calculation for general $\alpha$ (see Appendix
\ref{app:equi}). From Eq.\ (\ref{pathspatial}), it is clear that
the Ito convention with $\alpha =0$ is the simplest to deal with.
Another particular useful form of the path integral is obtained
using the Hubbard-Stratonovich transformation, which linearizes
the quadratic term
\begin{eqnarray}
P(x_f\,t_f| x_0\,t_0) = \int {\cal D} y \,\int_{x_0}^{x_f}{\cal
D}x\, \exp{-\int_{t_0}^{t_f}dt \, \left \{ {g(x)^2  \over 2  } \,
y^2(t) - \imath y(t) \left [ \partial_t x - f(x) +
\alpha\,g(x)\,g'(x)\right ] + \alpha f'(x) \right \}},
\label{path2}
\end{eqnarray}
where the measure now is $\int {\cal D} y \,\int_{x_0}^{x_f}{\cal
D}x = \int {d y_N \over 2 \pi }\ldots \int {d y_1  \over 2 \pi }
\int dx_{N-1}\ldots\int dx_{1}$. This result could, of course,
also have been obtained directly by substituting $k_i \equiv g
_i\,y_i$ in Eq.\ (\ref{conditional2}) and taking the continuum
limit. Note that in the discretized version of Eq.\ (\ref{path2}),
$y_n$ is associated with time $t^*_n = \alpha\, t_n +
(1+\alpha)\,t_{n-1}$.  This form of the path integral is closely
related to the MSR formalism \cite{MSR} to calculate response and
correlation functions.  This will be explored in Sec.\
\ref{subsec:correlation}.
\end{widetext}

It is interesting to see how the additional drift tern $f_1(x)$ in
the Langevin Equation [Eq.~(\ref{eq:f_1})] arises from the
constraints that equilibrium statistical mechanics impose on the
path integral formulation \cite{janssen}. Thermal systems must
obey detailed balance which states that
\bb
P(x_f\,t_f| x_0\,t_0) P_{eq}(x_0) = P(x_0\,t_f|
x_f\,t_0) P_{eq}(x_f).
\label{detail}
\en
The equilibrium distribution has the form $P_{eq}(x) = \exp{
-[\beta\, {\cal H}(x)]}$, and $P(x_0\,t_f| x_f\,t_0)$ is the
conditional probability for the reversed path, {\em i.e.}, for
$\bar{x}(t) =  x(-t)$. It turns out that the Stratonovich
convention is the simplest for the discussion of time-reversal
properties not only because it obeys the ordinary rule of
differential calculus, but also because it has the property that
the forward and backward paths are evaluated at the same points.
We will employ the Stratonovich convention below. First, we note
that
\begin{eqnarray}
{P_{eq}(x_f) \over P_{eq}(x_0)} &=&
\exp{ -[\beta{\cal H}(x_f)- \beta{\cal H}(x_0)] } \nonumber \\
&=& \exp{ -\left [ \int_{t_0}^{t_f} dt\,(\partial_t x )
\,{\partial \beta{\cal H}(x) \over \partial x}  \right ]},
\label{hamiltonian}
\end{eqnarray}
and that $P(x_0\,t_f| x_f\,t_0)$ can be obtained simply by noting
that the path associated with this distribution is the
time-reversal path of $P(x_f\,t_f| x_0\,t_0)$, which can be
written as
\begin{eqnarray} P(x_0\,t_f| x_f\,t_0) &=&
\int {\cal D} \tilde{y} \,\int_{x_0}^{x_f}{\cal D}x\, \label{reverse} \\
&\times& \,e^{-\int_{t_0}^{t_f}dt \, [ {g^2 \over 2  } \,
\tilde{y}^2 + \imath \tilde{y} \left [ \partial_t x + f - { 1\over
2} g g'\right ] + { 1 \over 2} f' ]}. \nonumber
\end{eqnarray}
Now, using Eqs.\ (\ref{detail}), (\ref{hamiltonian}), and
(\ref{reverse}) and comparing this term by term with exponential
in Eq.\ (\ref{path2}) [in the Stratonovich convention $\alpha
=1/2\,$], we see that
\begin{eqnarray}
\imath \tilde{y}(t) &=&
- \left [ \imath y(t) + {\partial \beta{\cal H}(x) \over \partial x}\right ], \\
f(x) &=& {1 \over 2}\,g(x) g'(x) - { g^2(x) \over 2}\,{\partial
\beta{\cal H}(x) \over \partial x}.
\label{f}
\end{eqnarray}
The first term in Eq.\ (\ref{f}) is identical to Eq.\ (\ref{fone})
in the Stratonovich interpretation. The second term is the
standard frictional term, from which we identify the dissipation
coefficient as $\Gamma(x) = \beta g^2(x)/2$, which is the Einstein
relation. This derivation again demonstrates that equilibrium
distribution is the only physics needed to fix $f(x)$ for a given
stochastic interpretation.

\subsection{Derivation of the Fokker-Planck Equation from the Path-integral}
\label{sub:derivation}

In this subsection, we derive the Fokker-Planck equation directly
from the conditional probability Eq.\ (\ref{conditional}), thereby
establishing the equivalence of the path integral formulation and
the Fokker-Planck equation for general $\alpha$. Let us rewrite
the conditional probability, Eq.\ (\ref{conditional2}), where we
set $x= x_i $, $t = t_i$, $x' = x_{i-1}$, $t'= t_{i-1}$, and $k_i
\equiv g_i\, y_i$:
\begin{eqnarray}
&&P(x\,t\,|x'\,t')  \hphantom{ppppppppppppppppppppppppppppppppp} \nonumber \\
&&= \int {d y_i \over 2 \pi }\,e^{-\Delta t \, [  {g_i^2  \over 2
} \, y_i^2 - \imath y_i \left [ { \Delta x_i \over \Delta t} - f_i
+ \alpha g_i g'_i\right ]
+ \alpha f'_i ]}, \nonumber\\
&&= \int {d y_i \over 2 \pi }\,e^{- \Delta t\,{\cal A}_i + \imath
y_i  \Delta x_i },
\end{eqnarray}
where\bb {\cal A}_i[x_i,x_{i-1};y_i] \equiv {g_i^2  \over 2  } \,
y_i^2 + \imath y_i \left [ f_i - \alpha g_i g'_i \right ] + \alpha
f'_i,
\label{A}
\en $\Delta x_i = x - x'$, and $\Delta t = t -t'$.  Our aim is to
calculate\bb P(x,t) = \int d x' \, P(x\,t\,|x'\,t')\,P(x' ,t'),
\label{cka}
\en to first order in $\Delta t$.  Expanding $P(x',t)$
\begin{eqnarray}
&& P (x', t') = P( x - \Delta x_i , t') \nonumber \\
&& \hphantom{pp} = P(x,t') - \Delta x_i\,{ \partial \over \partial
x }\,P(x,t') + { (\Delta x_i)^2  \over 2} { \partial^2 \over
\partial x^2} P(x,t'), \nonumber
\end{eqnarray}
and putting this back to Eq.\ (\ref{cka}), we find that it can
cast in the form\begin{eqnarray} { P(x,t)  - P(x ,t') \over \Delta
t}  &=&  \alpha(x)\,P(x,t')
+ \beta(x)\,{ \partial \over \partial x } P(x,t') \nonumber \\
&+& \gamma(x)\,{ \partial^2 \over \partial x^2 } P(x,t'),
\label{fpp}
\end{eqnarray}
where $\alpha(x) \equiv \lim_{\Delta t \rightarrow 0}\,{ \left
[\,{\cal I}_0(x) - 1 \,\right ] /\Delta t}$, $\beta(x) \equiv -
\lim_{\Delta t \rightarrow 0}\, { {\cal I}_1(x) / \Delta t}$, and
$\gamma(x) \equiv \lim_{\Delta t \rightarrow 0}\,{ {\cal I}_2(x)
/ (2\,\Delta t)}$.  The main task is to evaluate integral of the
form\bb {\cal I}_m(x) \equiv \int { dy_i \over 2 \pi} \int d
\Delta x_i \, (\Delta x_i)^{m}\,Q(x,\Delta x_i; y_i)\,e^{i y_i
\Delta x}, \en where\begin{eqnarray}
Q(x, \Delta x_i; y_i ) &=& e^{-\Delta t \,{\cal A}_i[x ,x';y_i] }, \nonumber \\
&=& \sum_k { (\Delta x_i)^k \over k!} \,\left. {\partial^k \over
\partial \Delta x_i^k }\, Q(x, \Delta x_i; y_i )\right |_{\Delta
x_i =0},
\end{eqnarray}
where in the last line, we have Taylor expanded the function $Q(x,
\Delta x_i; y_i )$. It is easy to see that \bb \int d \Delta x_i
\,(\Delta x_i)^{m}\,e^{i y_i \Delta x_i}  = 2 \pi (- \imath)^m {
\partial^m \over \partial y_i^m}\,\delta(y_i), \en and
therefore\bb {\cal I}_m(x) = \sum_k { (\imath)^{k+m} \over k!}
\left. { \partial^{m+k} \over \partial y_i^{m+k}}\, Q^{(k)}(x,0;
y_i)\right |_{y_i=0},
\label{q}
\en where $Q^{(k)}(x,0; y_i) \equiv  \left. {\partial^k Q(x,
\Delta x_i; y ) / \partial \Delta x_i^k } \right |_{\Delta x_i
=0}$. Using Eq.\ (\ref{q}), it is straightforward to compute
${\cal I}_m(x)$ to the first order in $\Delta t$; we
obtain\begin{eqnarray}
{\cal I}_0(x) &=& 1 - f'(x)\,\Delta t \nonumber \\
&+& (1-\alpha) \,\left \{ [g'(x)]^2 + g(x) g''(x)\right \}\,\Delta t,  \\
{\cal I}_1(x) &=& \left [\,f(x) - (2-\alpha) g(x) g'(x)\,\right ]\Delta t,  \\
{\cal I}_2(x) &=& g^2(x)\,\Delta t ,
\end{eqnarray}
with vanishing higher order terms,  i.e.\ ${\cal I}_n(x) = 0$ for
$n \geq 3$. Therefore, we have \begin{eqnarray}
\alpha(x) &=& - f'(x) + (1- \alpha) \left [ g'(x)^2 + g(x)g''(x) \right ],  \\
\beta(x) &=& - f(x) + (2- \alpha) g(x)g'(x), \\
\gamma(x) &=& { 1\over 2}\,g(x)^2.
\end{eqnarray}
This is equivalent to the Mori expansion \cite{risken}. It is
clear that with these coefficients Eq.\ (\ref{fpp}) becomes Eq.\
(\ref{fpe1}), the Fokker-Planck equation.

\subsection{Correlation, Response functions, and Fluctuation-Dissipation Theorem}
\label{subsec:correlation}

One of the advantages of the particular form of the path integral
in Eq.\ (\ref{path2}) is that correlation and response functions
can be computed conveniently from it. The average of any
functional ${\cal O}[x(t), y(t)]$ of $x(t)$ and $y(t)$ at fixed
$x_0$ is given by
\begin{eqnarray} \left \langle {\cal O}[x(t),
y(t)] \right \rangle_{x_0} = \int {\cal D} y \,\int_{x_0} {\cal
D}x\,{\cal O}[x(t), y(t)]\,e^{-{\cal S}}.
\end{eqnarray}
In particular, the two-point correlation function is
\bb
\langle\,x(t_1)\,x(t_2)\,\rangle_{x_0} = \int {\cal D} y
\,\int_{x_0} {\cal D} x\,x(t_1) x(t_2) \,e^{-{\cal S}},
\en
and the propagator function
\begin{eqnarray}
G(t_2,t_1) &\equiv& \langle\,x(t_2)\,[-\imath y(t_1)]\,\rangle_{x_0} \nonumber \\
&=& \int {\cal D} y \,\int_{x_0} {\cal D} x\,x(t_2) [-\imath
y(t_1)] \,e^{-{\cal S}}.
\label{response}
\end{eqnarray}
Physically, the propagator describes the response of the system to
a delta perturbation. One of the nice features of the propagator
function, which is useful in perturbative expansions, is that
causality is automatically built-in, {\em i.e.} $G(t,t') =0$ if $t
< t'$.  To see this, we go back to the discretized form of
the path integral and write $G(t,t')$ as\begin{eqnarray}
G(t_m,t_n) &=& \int
{d y_N \over 2 \pi }\ldots \int {d y_1  \over 2 \pi }
\int dx_{N}\ldots\int dx_{1} \, \nonumber \\
&\times& x_m\,[-\imath y_n]\,e^{-\sum_i \Delta t\,{\cal A}_i +
\sum_i \imath y_i  \Delta x_i },
\label{disresponse}
\end{eqnarray}
where ${\cal A}_i[x_i,x_{i-1};y_i]$ is defined in Eq.\ (\ref{A}).
First, let us consider $t_n > t_m$; each pair of the integrals
$\int {d y_i \over 2 \pi }\int dx_{i}$ in Eq.\ (\ref{disresponse})
gives $1$ for $i > n $. When integrating over $x_{n}$, we make use
the following identity
\begin{eqnarray}
\int dx_n\,Q[x_{n-1},\Delta x_n; y_n] e^{+ \imath y_n \Delta x_n}
\hphantom{ppppppppppp}\nonumber \\
= \sum_k  {(\imath)^k\,Q^{(k)}[x_{n-1},0,y_n] \over k!}
\,{\partial^k \over  \partial y_n^k}\, \delta(y_n),
\end{eqnarray}
which gives zero when integrating $y_n$. Thus, we have shown
$\langle x_m [-\imath y_{n}] \rangle_{x_0} = 0$ for all $n > m$.
Now, suppose $m=n$, one can show that using the above identity,
$\langle x_n [ -\imath y_n] \rangle_{x_0} = 1$.  Clearly,
$\langle x_m [ -\imath y_{n} ] \rangle_{x_0} \neq 0$, if $m > n$.
Thus, we have shown how the path integral enforces causality, {\em
i.e.\ } $G(t, t' ) = 0$ if $t< t'$, and $G(t, t' ) \neq 0$ if $t >
t'$. However, there is a subtle point about the value of $G(t,t)$
in the continuum limit, which has to be consistent with the
$\alpha$-convention.  The simplest way do this is to note that
since $y_n$ is really associated with time at $t^*_n =
\alpha\,t_{n} + (1- \alpha)\,t_{n-1}$, we have to
evaluate\begin{eqnarray}
G(t,t) &\equiv& \langle\,x(t^*_n) [ - \imath y_n] \rangle_{x_0} \nonumber \\
&=& \langle\,[\,\alpha\,x_{n} + (1-\alpha)\,x_{n-1}] ( - \imath
y_n)\,\rangle_{x_0} = \alpha. \nonumber
\end{eqnarray}

Now, we specialize to a system near equilibrium, and we
investigate how the path integral describes properties such as the
Fluctuation-Dissipation Theorem.  The equilibrium average of any
function ${\cal O}[x(t), y(t)] $ of $x(t)$ and $y(t)$ is defined
as
\bb
\left \langle {\cal O}[x(t), y(t)] \right \rangle_{eq} = \int
{\cal D} y \,\int {\cal D}x\,{\cal O}[x(t), y(t)]\,e^{-{\cal S}}
P_{eq}(x_0).
\en
Note that equilibrium averages are independent of
$\alpha$, provided that we add the additional drift $f_1(x)$. When
the system is under a time-dependent physical force $h(t)$, the
total Hamiltonian is ${\cal H}_T = {\cal H}_0(x) - x(t) h(t),$ so
that
\begin{eqnarray} f(x,t) &=& (1-\alpha) g g' - \Gamma(x){\partial
{\cal H}_T \over \partial x}
= f_0(x) + \Gamma(x) h(t), \nonumber \\
f_0(x) &=& ( 1- \alpha ) g g' - \Gamma(x){\partial {\cal H}_0
\over \partial x}. \nonumber
\end{eqnarray}
Therefore, we have
\begin{eqnarray} \left. { \delta \langle x(t)
\rangle \over \delta h(t')}\right |_{h(t)=0} &=& { 1 \over 2 k_B T
}\left \langle\,x(t) \left \{ -\imath y(t')\,g[x(t')]^2 \right \}
\right \rangle_{eq} \nonumber \\
&-&  { \alpha  \over  k_B T }
\left \langle\,x(t) g[x(t')]g'[x(t')]\right \rangle_{eq}
\label{physresponse}\\
&\equiv& {\cal \chi}_{xx}(t,t'). \nonumber
\end{eqnarray}
We observe that the response $\chi_{xx}(t,t')$ to a physical
forces and the propagator $G(t,t')$ defined in
Eq.~(\ref{response}) are different, although they are proportional
to each other for the case of uniform diffusion constant. In
particular, there is an additional term arising from the
normalization factor $f'$ in the action and it is absent when the
diffusion constant is spatially uniform.  By integration by parts,
the first term in the bracket can be evaluated to
be\begin{eqnarray} \langle\,x(t) \left \{ -\imath
y(t')\,g[x(t')]^2 \right \} \rangle_{eq}
\hphantom{\partial_{t'} x(t') - f_0[x(t')]\partial_{t'} x(t') ++} \nonumber \\
= \langle\,x(t) \left \{ \partial_{t'} x(t') - f_0[x(t')] + \alpha
g[x(t')]g'[x(t')] \right \} \rangle_{eq}. \nonumber
\end{eqnarray}
Therefore, the physical response function is
\bb
{\cal\chi}_{xx}(t,t')  = { 1 \over 2 k_B T }\, \left \langle\,x(t)
\left \{\partial_{t'} x(t') - {\cal B}[x(t')] \right \} \right
\rangle_{eq},
\label{responsex}
\en
where ${\cal B}(x) \equiv gg' - \Gamma(x){ \partial {\cal H}_0
/ \partial x}$. Note that the physical response function is
independent of $\alpha$, as it should be; note also the a drift
proportional to $gg'$ arises from spatial varying diffusion
constant.  To proceed further, we note that as a consequence of
the detailed balance condition, Eq.\ (\ref{detail}),  the
equilibrium correlation function is symmetric with respect to
exchange of $t \leftrightarrow t'$:
\begin{eqnarray}
\left \langle
{\cal O}_1[x(t)] {\cal O}_2[x(t')] \right \rangle_{eq} = \left
\langle {\cal O}_1[x(t')] {\cal O}_2[x(t)] \right \rangle_{eq}.
\nonumber
\end{eqnarray}
Applying this to Eq.\ (\ref{responsex}) and subtracting the
results, we have
\bb \left [ \partial_t - \partial_{t'} \right ]
\left \langle x(t) x(t') \right \rangle_{eq} = 2 k_B T \left [
{\cal \chi}_{xx}(t',t) - {\cal \chi}_{xx}(t,t') \right ].
\nonumber
\en
Since the correlation function is time translational
invariant, we must have $ \partial_{t'} \left \langle x(t) x(t')
\right \rangle_{eq} = -
\partial_{t} \left \langle x(t) x(t') \right \rangle_{eq}$.
Thus,
\bb
\partial_t \left \langle x(t) x(t') \right \rangle_{eq} = - k_B T\,
\left [ {\cal \chi}_{xx}(t,t') - {\cal \chi}_{xx}(t',t) \right ].
\nonumber
\en
This is the Fluctuation-Dissipation Theorem. To put
it in a more traditional form, we note that ${\cal
\chi}_{xx}(t',t) = 0$ when $t > t'$, and we can write
\bb
{\cal
\chi}_{xx}(t - t') = - { 1 \over k_B T }\,
\partial_t \left \langle x(t) x(t') \right \rangle_{eq}\,\theta(t-t'),
\nonumber
\en
where $\theta(t)$ is the Heaviside unit step function.  The Fourier transform of the
response function is given by
\bb \left \langle
x(\omega) x(-\omega) \right \rangle_{eq} = {2 k_B T \over
\omega}\,\mbox{Im}\, {\cal \chi}_{xx}(\omega),
\label{fdt}
\en
which is of the form that is commonly quoted in the literature.

\subsection{Perturbation Theory}
\label{subsec:perturbation}

One of the advantages of the path integral formulation of
stochastic dynamics is that it is by construction a field theory
that facilitates systematic perturbative calculation of
correlation functions.  In particular, for systems with
state-dependent dissipative coefficients, the resulting Langevin
equation is generally nonlinear, and perturbation theory is a
convenient way to derive the mode-coupling theory \cite{reichman}.
Thus, in this subsection, we set up the perturbation theory for a
systematic calculation of correlation and response functions in
the deviation of the diffusion coefficient from spatial
uniformity. First, we need to set up the generating functional.
Note that $P_{eq}(x_0)$ satisfies
\bb
P_{eq}(x) = \int dx_{s}\,P(x\,t\,|x_s\,t_s)
P_{eq}(x_s),
\en
which implies that
\bb
P_{eq}(x)= \lim_{t_s
\rightarrow -\infty}\, P(x\,t\,| x_s\,t_s).
\en
Thus, the equilibrium averages can be written as
\begin{eqnarray}
\left \langle {\cal O}[x(t), y(t)] \right \rangle_{eq} &=&
\int {\cal D} y \,\int {\cal D}x\,
{\cal O}[x(t), y(t)]\,e^{-{\cal S}} P_{eq}(x_0) \nonumber \\
&=& \int {\cal D} y \,\int {\cal D}x\,{\cal O}[x(t),
y(t)]\,e^{-{\cal S}}, \nonumber
\end{eqnarray}
where in the last line, the limit of the time integration in the
action ${\cal S}$ is extended to $-\infty$ to $\infty$. This
allows us to define the generating function for equilibrium
averages by
\bb {
\cal Z}[F, \tilde{F}] = \int {\cal D} y \,\int
{\cal D}x\,e^{-{\cal S} + \int dt [ x(t) F(t) -\imath y(t)
\tilde{F}(t) ] }.
\en
The correlation functions and the propagator are simply functional
derivatives of ${\cal Z}$.  This sets up the MSR perturbation
scheme \cite{MSR} that allows the immediate application of all of
the powerful techniques of field theory, including the
renormalization group, to nonlinear stochastic problems. It is
customary to introduce the variables $\xh(t) \equiv -\imath y(t)$.
Note that in the perturbation expansion, all the $\alpha$
dependent terms cancel provided that we use $\langle x(t)
\xh(t)\rangle_0 = \alpha$ (see Appendix \ref{app:perturbation}).
Therefore, it is convenient to use $\alpha =0$ at the outset.

As an informative model calculation, we explore the problem in
which a particle diffusing with $D(x) = D_0 \left ( 1 +
\gamma\,x^2 \right )$ confined in a harmonic potential, ${\cal H}
= k x^2 /2$.  If the confining potential is turned off,
this problem is exactly solvable, as shown in Sec.\
\ref{subsec:wierd}. It can also be solved exactly when $\gamma =0$
but not when $\gamma \neq 0$, and a perturbative expansion
in $\gamma$ is useful. The goal of this exercise is to compute the
propagator $\langle x \xh \rangle$ and the correlation function
$\langle x x \rangle$ separately and to check that the Fluctuation-Dissipation theorem is
satisfied. For simplicity, we work in the Ito convention (see
Appendix \ref{app:perturbation} for general $\alpha$), and set
$k_B =1$. According to the formalism, we have
\begin{eqnarray}
f(x) &=& g(x) g'(x) - \Gamma(x) { \partial {\cal H} \over \partial x}
\nonumber \\
&=& - \Gamma_0 ( k - 2 \gamma T )\,x - \Gamma_0 k \gamma x^3.
\label{fx}
\end{eqnarray}
It should be pointed out that from Eq.\ (\ref{fx}) one might at
first sight conclude that there is a broken-symmetry state, when
$k < 2 \gamma T $, with $\langle x \rangle \neq 0$. But we know
that this cannot happen because the stationary distribution is in
fact the Boltzmann distribution.  Therefore, one could get the
wrong physics if one only looks at ``classical" trajectory, {\em
i.e.} solution to $\partial_t x = f(x)$, which maximizes the
action ${\cal S}$ in the Ito convention.  This shows again the
importance of noise in these problems.

\begin{figure}
\resizebox{2in}{!}{\includegraphics{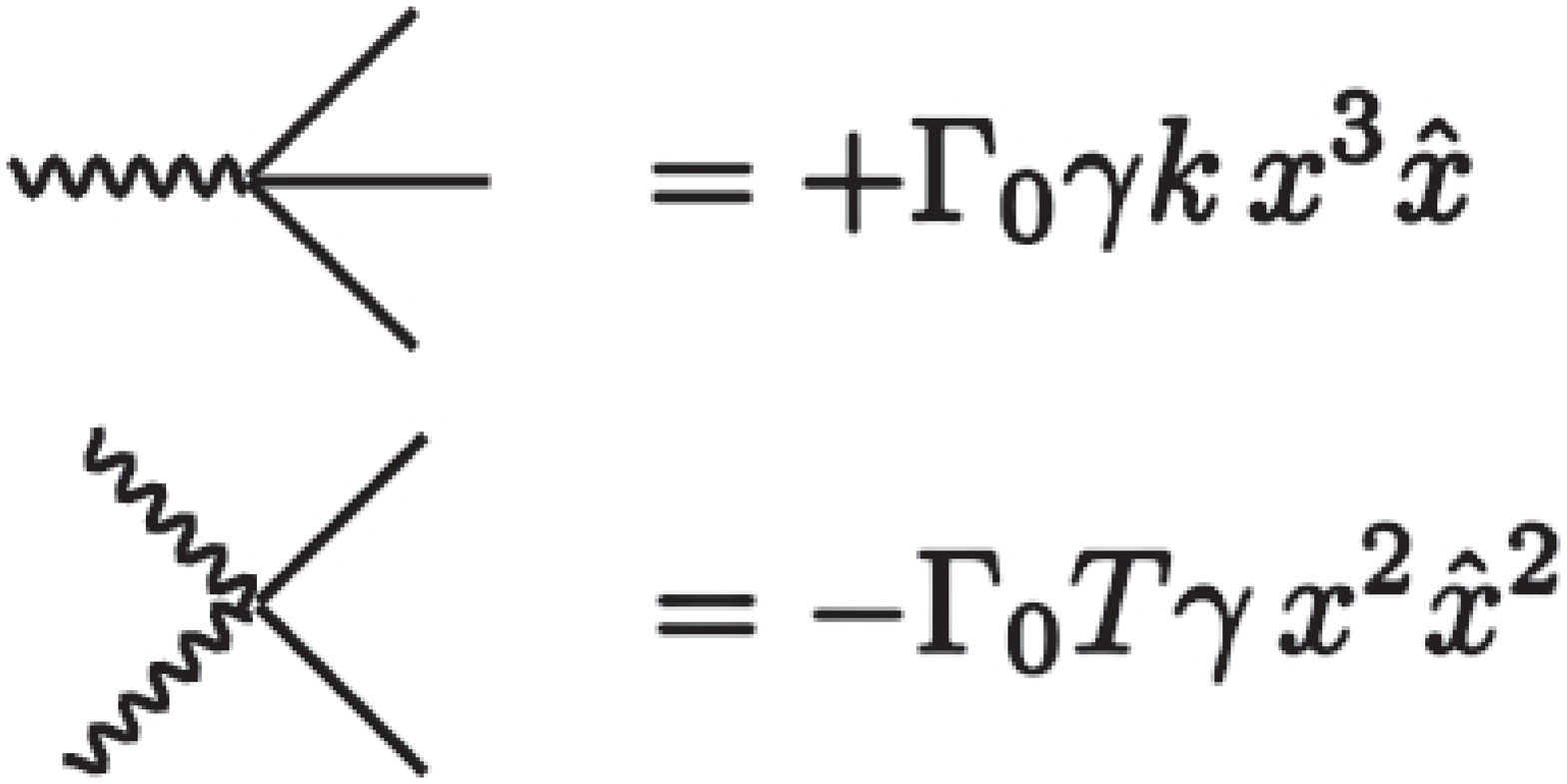}}
\caption{\label{fig:vertices} The two vertices corresponding to Eq.\ (\ref{actioni}) for
a particle diffusing in a spatially varying diffusion coefficient given by $D(x) = D_0 ( 1+ \gamma\,x^2)$,
and confined in a harmonic potential.}
\end{figure}

\begin{figure}
\resizebox{2.5in}{!}{\includegraphics{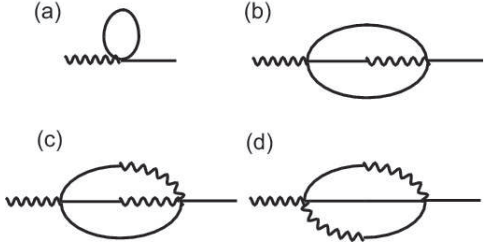}}
\caption{\label{fig:diagrams} Diagrams that contribute to the self-energy $\Sigma(\omega)$. Note that
diagram $d$ is identically zero.}
\end{figure}

The unperturbed and perturbing actions are
\begin{eqnarray}
{\cal S}_0 &=& \int dt \left [ -D_0 \xh^2 + \xh
\left ( \partial_t x + \Gamma_0 k' x \right ) \right ] \\
{\cal S}_I &=&  \Gamma_0 \gamma \int dt \left [ k\,x^3 \xh -
T\,x^2 \xh^2 \right ],
\label{actioni}
\end{eqnarray}
where $k' \equiv k - 2\gamma T $.  Introducing the state vector
$x_{\alpha} = ( \xh,x)$, we can write ${\cal S}_0$ as
\bb
{\cal
S}_0 = { 1\over 2} \int { d\omega \over 2 \pi} x_{\alpha}(\omega)
G^{-1}_{\alpha \beta}(\omega) x_{\beta}(-\omega), \en where\bb
G^{0\,-1}_{\alpha \beta}(\omega) = \left(
                                  \begin{array}{cc}
                                    -2 D_0  &  -\imath \omega + \Gamma_0 k' \\
                                    \imath \omega +  \Gamma_0 k'  & 0\\
                                  \end{array}
                                \right),
\nonumber
\en
and thus
\bb G^{0}_{\alpha \beta}(\omega) = \left(
                                  \begin{array}{cc}
                                    0  &  { 1\over \imath \omega + \Gamma_0 k' }\\
  { 1\over -\imath \omega +  \Gamma_0 k'}   & { 2 D_0 \over | - \imath \omega + \Gamma_0 k'|^2} \\
                                  \end{array}
                                \right),
\nonumber
\en
from which we can read off the bare propagator and
the zeroth-order correlation function:
\begin{eqnarray}
\langle \xh(\omega) \xh(\omega') \rangle_0 &=& 0  \nonumber \\
\langle \xh(\omega) x(\omega') \rangle_0 &=& {
 1\over \imath \omega + \Gamma_0 k' }\,\delta(\omega +\omega')\nonumber \\
\langle x(\omega) \xh(\omega')  \rangle_0 &=&
{ 1\over - \imath \omega + \Gamma_0 k' }\,\delta(\omega +\omega')\nonumber \\
\langle x(\omega) x(\omega') \rangle_0 &=& { 2 D_0 \over | -
\imath \omega + \Gamma_0 k'|^2}\,\delta(\omega +\omega').
\nonumber
\end{eqnarray}
The interacting ${\cal S}_I$ consists of two vertices that are depicted in
Fig.\ \ref{fig:vertices}.  To second order in $\gamma$, the inverse of the propagator,
$G^{-1}(\omega)$, can be written in frequency space as
\bb
G^{-1}(\omega) = -\imath\,\omega + \Gamma_0 ( k+ \gamma T)  +
\Sigma(\omega).
\en
The self-energy $\Sigma(\omega)$ are computed
from the diagrams listed in Fig.\ \ref{fig:diagrams} and it is
given by
\begin{eqnarray}
\Sigma(\omega) &=& 6\, k\,(\Gamma_0
\gamma)^2\,\left [\,{ 2\,T} A(\omega) - { 3\,k \,}
B(\omega)\,\right ], \nonumber
\end{eqnarray}
where
\begin{eqnarray}
A(\omega) &=& \int {d\omega_1 \over 2
\pi}\,\int {d\omega_2 \over 2 \pi}\,
G_0(\omega -\omega_1-\omega_2) G_0(\omega_1) C_0(\omega_2),\nonumber \\
B(\omega) &=& \int {d\omega_1 \over 2 \pi}\,\int {d\omega_2 \over
2 \pi}\, G_0(\omega -\omega_1-\omega_2) C_0(\omega_1)
C_0(\omega_2),\nonumber
\end{eqnarray}
where $C_0(\omega) = \langle x(\omega) x(-\omega) \rangle_0$ is
the zero-order correlation function.  After some algebra, we
obtain
\bb
G^{-1}(\omega) = - \imath \omega  +  \Gamma_0 k \left
[\, 1 + {\gamma T / k} - 6\,(\gamma T /k)^2 \mu(\omega/\Gamma_0 k)
\,\right ], \nonumber
\en
where $\mu(s) \equiv (- \imath s +3)^{-1}$.

\begin{figure}
    \resizebox{2.5in}{!}{\includegraphics{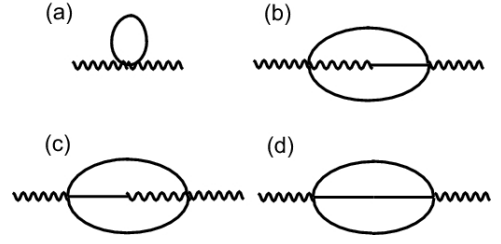}}
    \caption{\label{fig:diagramn} Diagrams that contribute to the noise $D(\omega)$.}
\end{figure}

\begin{figure}
    \resizebox{2.7in}{!}{\includegraphics{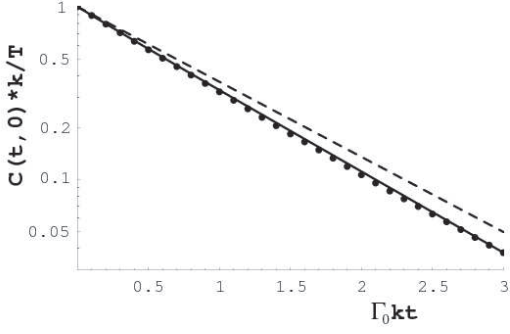}}
    \caption{\label{fig:corrfun}  A plot of the correlation function of $x$, $C(t,0)$, as a function
    of $t$ for $\gamma T /k = 0.15$.  The solid line represents $C(t,0)$ as given by Eq.\ (\ref{corrfunt}),
    which is calculated from the perturbation theory to second order in $\gamma T /k $.  The data points
    are obtained from numerical simulation of the corresponding Langevin equation.  Clearly, the result from
    the perturbation theory agree excellently with the simulation.  The dashed line represents the bare (zeroth
    order) correlation function, which has a decaying rate that is different from the case if
    the spatial-varying dissipative coefficient.}
\end{figure}

Next, we turn to the correlation function $C(t,t') =
\langle\,x(t) x(t')\,\rangle$, which can be written in the form:
$C(\omega) = 2 D(\omega) \left | G(\omega)\right |^2 $, with
$D(\omega)$ are computed from diagrams listed in Fig.\
\ref{fig:diagramn} and it is given by
\begin{eqnarray}
D(\omega)
&=& D_0 \left ( 1 + {\gamma T \over k} \right )
+  k \Gamma_0^2 \gamma^2 \left [\,3 k E(\omega) - 8 T H(\omega)\,\right ]\nonumber  \\
E(\omega) &=& \int {d\omega_1 \over 2 \pi}\,\int {d\omega_2 \over
2 \pi}\,C_0(\omega -\omega_1 -\omega_2)
C_0(\omega_1) C_0(\omega_2), \nonumber  \\
H(\omega) &=& \int {d\omega_1 \over 2 \pi}\,\int {d\omega_2 \over 2 \pi}\, C_0(\omega_1) C_0(\omega_2) \nonumber \\
&\times& \left [\, G_0(\omega -\omega_1 - \omega_2) + G_0(
\omega_1 + \omega_2 -\omega  )\,\right ]. \nonumber
\end{eqnarray}
After some algebra, we find
\bb
D(\omega) = 2D_0 \left [\,1 +
(\gamma T /k) - 6 (\gamma T / k)^2\,\mbox{Re}\,\mu(\omega/\Gamma_0
k)\,\right ], \nonumber
\en
and we evaluate the correlation function,
\begin{eqnarray}
C(t,0) &=& { T\over k}\,\left [ {
3 \alpha_2 - \alpha_1\,\left ( 1 + {\gamma T \over k} \right )
\over \alpha_2^2 - \alpha_1^2 }\,\right ] e^{- \alpha_1 t}  \nonumber \\
&+& { T\over k}\, \left [ { \alpha_2\, \left ( 1 + {\gamma T\over k} \right ) -
3\,\alpha_1 \over \alpha_2^2 - \alpha_1^2 }\,\right ] e^{- \alpha_2 t},
\label{corrfunt}
\end{eqnarray}
with decay rates
\begin{eqnarray}
\alpha_{1} &\equiv&  {
\Gamma_0 k \over 2}
\left [\, 4+ \gamma T /k\,- \sqrt{\,4 - 4 \gamma T/k + 25\,(\gamma T/k)^2\,}\,\right ], \nonumber \\
&\approx& { \Gamma_0 k } \left [ 1 +  {\gamma T/ k}
- 3 \left ({\gamma T/ k} \right )^2 \right ],\,\,\,\,\mbox{for $\gamma T/k \ll 1$},\nonumber\\
\alpha_{2 } &\equiv& { \Gamma_0 k \over 2} \left [\, 4+ \gamma T
/k\,+\sqrt{\,4 - 4 \gamma T/k + 25\,(\gamma T/k)^2\,}\,\right ],
\nonumber \\
&\approx& 3 \Gamma_0 k \left [  1+ (\gamma T/k)^2 \right ],\,\,\,\,\mbox{for $\gamma T/k \ll 1$}. \nonumber
\end{eqnarray}
Note that there are now two decaying modes with a fast mode
$\alpha_2$ and a slow mode $\alpha_{1}$ in the system in contrast
to the case with with uniform diffusion. Note also that $\langle
x^2(0) \rangle = T/k$ as it should be.  If we did not put in the
extra drift term $gg'$ in $f(x)$, this relation would not hold.
In fact, it would have been $\langle x^2(0) \rangle = T/(k-\gamma
T)$, which violates the equipartition theorem. This is yet another
demonstration that this extra drift term $gg'$ is needed to ensure
the correct thermodynamic properties. In Fig.\ \ref{fig:corrfun},
we plot the correlation function in Eq.\ (\ref{corrfunt}) and the
numerical simulation of the Langevin equation describing this
system for $\gamma T/ k = 0.15$. Clearly, the second order
perturbation theory agrees very well with the simulation.  Note,
however, that when $\gamma T/k \sim 1$, $\alpha_1$ becomes
negative, signalling the breakdown of perturbation theory.

\begin{figure}
    \resizebox{2.5in}{!}{\includegraphics{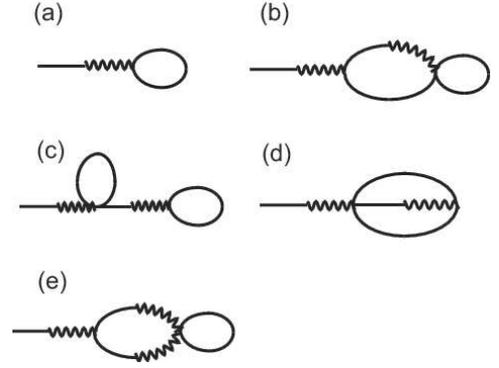}}

    \caption{\label{fig:diagramr} Diagrams that contribute to the physical response function $\chi_{xx}(\omega)$.}
\end{figure}

Finally, we demonstrate FDT to second order in perturbation
theory. The physical response function is given by
\begin{eqnarray}
\chi_{xx}(t, t') &=&  \Gamma_0 \left \langle\,x(t)\,\xh(t')
 \left [\, 1+ \gamma\,x^2(t')\,\right ] e^{-{\cal S}_I} \right \rangle_0  \nonumber \\
 &=& \Gamma_0 G(t, t') + \Gamma_0 \gamma G_0(t,t') C_0(0)  \nonumber \\
 &-&
 \Gamma_0 \gamma \left \langle\,x(t)\,\xh(t') x^2(t') {\cal S}_I \right \rangle_0,
 \nonumber
\end{eqnarray}
which corresponds to the diagrams in Fig.\ \ref{fig:diagramr}.  We
find
\bb
\chi_{xx}(\omega)  = \Gamma_0 G(\omega) \left [ 1 +
\gamma T/k - 6\, (\gamma T /k)^2 \mu(\omega/\Gamma_0 k) \right ].
\label{physical}
\en
Note that this clearly shows that the physical response
function and the propagator are different. Taking the imaginary
part of Eq.\ (\ref{physical}), it can be easily verified that the
Fluctuation-dissipative theorem Eq.\ (\ref{fdt}) is indeed
satisfied to second order in perturbation theory.

\section{N-Components Langevin Equation}
\label{sec:higher}

Many physical problems involve more than one variable and some of
the issues we have addressed so far may not apply to
higher-dimensional systems.  For example, the drift term $f(x)$ in
1-D can always be written as a derivative of another function,
{\em i.e.} 1-D systems are conservative, however, for higher
dimensional systems, this may not be true.  A complete analysis of
higher dimensional systems requires a separate publication.  Here,
we briefly discuss the Fokker-Planck equation and the path
integral in the $\alpha$-convention for a multidimensional
Langevin equation of the form
\bb
\partial_t x_i = f_i(x_1, \cdots, x_N) + g_{ij}(x_1, \cdots, x_N) \eta_j(t),
\label{langN}
\en
where $\eta_i(t)$ are the noises, with zero mean and
correlation given by
\bb
\langle\,\eta_i(t)\eta_j(t')\,\rangle =
\delta_{ij} \delta(t-t').
\en
In Eq.\ (\ref{langN}) and the following, Einstein summation is assumed. We focus on the case where the
system is near thermal equilibrium and address, as we did in the
1D case, how the Boltzmann distribution determines the form of
$f_i$ and $g_{ij}$ in the $\alpha$-convention. The Fokker-Planck
equation corresponds to Eq.\ (\ref{langN}) can be derived
following the same procedure as outlined in Sec.\
\ref{subsec:fpe}. In the $\alpha$-convention, we find
\bb
{\partial
{\cal P} \over \partial t} = { \partial \over \partial x_i} \left
\{ - \left [ f_i + \alpha { \partial g_{il} \over \partial x_k }\,
g_{kl}  \right ]\, {\cal P} + { 1 \over 2 } {\partial \over
\partial x_j} \left ( g_{il}\,g_{jl} {\cal P}\right ) \right \},
\label{fpN}
\en
where ${\cal P}[ \{x \}, t]$ is the joint probability distribution
of $x_i$ at time $t$. If there are only dissipative terms and no
reactive terms in $f_i$, the constraint that ${\cal P}$ reach a
long-time state of thermal equilibrium value proportional to $\exp
- \beta {\cal H}$ requires that
\bb
f_i(\{x\}) = { 1 \over 2} { \partial
\over \partial x_j} \left ( g_{ik}\,g_{jk}\right ) - \alpha {
\partial g_{ij} \over \partial x_l}g_{lj}- { 1\over 2} \beta
g_{ik}g_{jk}
 {\partial {\cal H} \over \partial x_j},
\en
in order that Eq.\ (\ref{fpN}) reduce to
\bb
{\partial {\cal
P} \over \partial t} = { \partial \over \partial x_i} \left ( { 1
\over 2 }  g_{il}\,g_{jl} \right ) \left [  \beta {\partial {\cal
H} \over \partial x_j}\,{\cal P} + {\partial  {\cal P} \over
\partial x_j} \right ] ,
\en
with the steady-state solution ${\cal P} \propto \exp - \beta
{\cal H}$.

The diffusion matrix is defined as
\bb
D_{ij}(\{x\}) = { 1 \over 2 }  g_{il}\,g_{jl} = k_B T
\Gamma_{ij}(\{x\}),
\en
where $\Gamma_{ij}(\{x\})$ is the matrix of dissipative
coefficients.  Thus
\bb
f_i(\{x\}) =  { \partial
D_{ij}(\{x\}) \over \partial x_j} - \alpha { \partial g_{ij} \over
\partial x_l}g_{lj}- \Gamma_{ij}(\{x\}) {\partial {\cal H} \over
\partial x_j},
\en
Note that since the diffusion matrix $D_{ij}(\{x\})$ is a
symmetric with respect to $i \leftrightarrow j$, it has only
$N(N+1)/2$ independent entries, and we may impose $N^2 - N(N+1)/2
= N(N-1)/2$ constraints on $g_{ij}(\{x\})$ without sacrificing the
physical content.  We could, for example chose $g_{ij}$ to be
symmetric in which case, it is simply the square root of $k_B T
\Gamma_{ij}$

To derive the path integral, we first discretize Eq.\
(\ref{langN}) as
\bb x_i^{(n)} = x_i^{(n-1)} + \Delta t f_i^{(n)} +
g_{ij}^{(n)} \int_{t_{n-1}}^{t_n} ds\,\eta_j(s),
\en
and introduce
\begin{widetext}
\begin{eqnarray} h_i\left [ \left \{x_k^{(n)}
\right \}, \left \{x_k^{(n-1)} \right \} \right ] \equiv
[g^{(n)}]_{ij}^{-1}  \left [ x_j^{(n)} - x_j^{(n-1)} - \Delta t
f_j^{(n)} \right ] - \int_{t_{n-1}}^{t_n} ds\,\eta_i(s),
\end{eqnarray}
where $f_i^{(n)}= f_i( \{ \alpha x_i^{(n)} + (1-\alpha )
x_i^{(n-1)} \})$ and $g_{ij}^{(n)}= g_{ij}( \{ \alpha x_i^{(n)} +
(1-\alpha ) x_i^{(n-1)} \})$. Following the basic steps as
outlined in Sec.\ \ref{sec:pathintegral}, we can write the
conditional probability as
\begin{eqnarray}
{\cal P}(
\{x_k^{(n)}\}\,t_n | \{x_k^{(n-1)} \}\,t_{n-1} ) =
 \det \left | { \partial h_i \over \,\partial x_k^{(n)} } \right | \left \langle \prod_{i=1}^N \,
\delta \left [ h_i( \{x_k^{(n)} \}, \{x_k^{(n-1)}  \} )\right ]
\right \rangle.
\end{eqnarray}
Taking the derivative of $h_i$ explicitly, we find
\begin{eqnarray}
{ \partial h_i \over \,\partial x_k^{(n)} }  = g^{-1}_{il} \left
(\,\delta_{lk} - {\cal M}_{lk}\,\right ), \nonumber
\end{eqnarray}
where we have defined the matrix ${\cal M}_{lk}$
by
\begin{eqnarray} {\cal M}_{lk} &\equiv& \alpha \Delta t {
\partial f_l\over \partial x_k} + \alpha {\partial g_{lm} \over
\partial x_k }\,g^{-1}_{ij}\left [\,x_j^{(n)} - x_j^{(n-1)} -
\Delta t f_j^{(n)} \right ].
\end{eqnarray}
Using the identity $\det \hat{X} = \exp \mbox{Tr} \hat{X}$, the
determinant can be evaluated to give
\begin{eqnarray}
\det \left |
{ \partial h_i \over \,\partial x_k^{(n)} } \right | = { 1 \over
\det g_{ij}} \left [\, 1 - {\cal M}_{ll} +  {1 \over 2} \,\left (
{\cal M}_{ll}{\cal M}_{kk}- {\cal M}_{lk}{\cal M}_{kl} \right ) +
\cdots \right ].
\end{eqnarray}
Therefore, the conditional probability can be written as
\bb {\cal
P}( \{x_k^{(n)} \}\,t_n | \{x_k^{(n-1)} \}\,t_{n-1} ) =
\prod_{i=1}^N\, \int {d k_i \over 2 \pi \det g_{ij}} \left [ 1 -
{\cal M}_{ll} + {1 \over 2} \,\left ( {\cal M}_{ll}{\cal M}_{kk}-
{\cal M}_{lk}{\cal M}_{kl} \right ) \right ] e^{ \imath k_i
g_{ij}^{-1} \left ( \Delta x_j - \Delta t f_j \right ) - { 1\over
2} k_i^2 \Delta t },
\en
where we have only kept terms up to order
$\Delta t$.  Following the similar procedure leading to Eq.\
(\ref{conditional}) for the 1-D case, we find
\begin{eqnarray}
{\cal P}( \{x_k^{(n)} \}\,t_n | \{x_k^{(n-1)} \}\,t_{n-1} )  &=&
\prod_{i=1}^N\, \int {d k_i \over 2 \pi \det g_{ij}} \left [  1 -
\alpha \Delta t \partial_l f_l + \alpha \imath k_m
\partial_l g_{lm}  \Delta t + {\alpha^2 \over 2} \Delta t \left (
\partial_k g_{lm} \partial_l g_{km} - \partial_l g_{lm} \partial_k g_{km}\right ) \right ] \nonumber \\
&\times& e^{ \imath k_i g_{ij}^{-1} \left ( \Delta x_j - \Delta t
f_j \right ) - { 1\over 2} k_i^2 \Delta t },
\label{multi} \\
&=& \int {d y_k \over 2 \pi}\, e^{+ \imath y_k ( \Delta x_k - f_k
\Delta t + \alpha\,g_{kj} \partial_l g_{lj} \Delta t ) - {1\over
2} g_{lk} g_{kj} y_l y_j - \alpha \Delta t \partial_l f_{l}  - {
\alpha^2 \over 2} \left ( \partial_k g_{lm} \partial_l g_{km} -
\partial_l g_{lm} \partial_k g_{km}\right ) \Delta t}, \nonumber
\end{eqnarray}
where in the last line, we have exponentiate terms in the bracket
and substituted $y_i = g_{ij} k_j$. In the continuum limit, we
have
\begin{eqnarray}
&& {\cal P}( \{x_k^{(f)} \}\,t_f|
\{x_k^{(0)} \}\,t_{0} ) = \int_{x_k^{(0)}}^{x_k^{(f)}}{\cal D}x_k
\int {\cal D}y_k \, e^{-{\cal S}},
\nonumber \\
&& {\cal S} = \int dt \left [ {1\over 2}\,g_{lk} g_{jk} y_l y_j -
\imath y_k \left (\,\partial_t x_k - f_k + \alpha\,g_{kj}
\partial_l g_{lj} \, \right ) + \alpha\,\partial_l f_{l} + {
\alpha^2 \over 2}\,\left (\,\partial_k g_{lm} \partial_l g_{km} -
\partial_l g_{lm} \partial_k g_{km} \right )\right ].
\end{eqnarray}

Note that there is an extra term proportional to $\alpha^2$. This
term is identically zero for 1-D system.  If we start with Eq.\
(\ref{multi}) and follow the procedure as outlined in Sec.\
\ref{sub:derivation}, we can show that the path integral is
equivalent to the Fokker-Planck equation in Eq.\ (\ref{fpN}).
\end{widetext}

\section{Conclusion}
\label{sec:conclusion}

In this paper, we have examined a thermodynamically consistent
Langevin formulation of the Brownian motion with a diffusion
coefficient that depends on space.  We argue, in particular, that
the requirement that the Boltzmann distribution be reached in
equilibrium determines the interpretation of stochastic integrals
arising from multiplicative noise in the Langevin equation. We
hope that this paper clarifies some of the confusion over these
stochastic issues that have persisted for some time.  We have also
constructed path integral representations of the
Langevin equations with multiplicative noise, and we used this
representation as a starting point for the development of a
systematic perturbation theory.  Such a formulation can be
employed to treat nonlinear stochastic equations arising from a
variety of problems. Future work includes generalizing this
formulism to ``fields" and examines how state-dependent
dissipative coefficients may give rise to long-time tails and
corrections to scaling in dynamic critical phenomena. Of course,
one of the most interesting open questions is whether there is an
equivalent criteria for systems that are driven far from
equilibrium.

\begin{acknowledgments}
This work was supported in by by US National Science Foundation under Grant No. DMR 04-04670.
\end{acknowledgments}

\appendix

\section{Connection between $\alpha$ and $\theta(0)$}
\label{app:alternative}

In this Appendix, we outline the connection between $\alpha$ and
the Heaviside unit step function, $\theta(t)$ evaluated at
$t=0$.  For simplicity, we set $f(x)=0$ in Eq.\ (\ref{spatial}):
\bb
\partial_t x = g(x) \eta(t)\label{geta}.
\en
Using the $\alpha$-convention rule Eq.\ (\ref{alpha}), we have
for $\Delta x(t) \equiv x(t+\Delta t) - x(t)$
\begin{eqnarray}
\Delta x(t) &=& g[x(t) + \alpha \Delta x] \int_{t}^{t +\Delta t} ds\,\eta(s) \nonumber \\
&=& g[x(t)] \int_{t}^{t +\Delta t} ds\,\eta(s) \nonumber \\
&\phantom{+}& + \,\alpha\,g'[x(t)]\,\Delta x(t)\, \int_{t}^{t +\Delta t} ds\,\eta(s) + \cdots \nonumber \\
&=& g[x(t)] \int_{t}^{t +\Delta t} ds\,\eta(s) \nonumber \\
&\phantom{+}& +\,  \alpha g g' \int_{t}^{t +\Delta t}
ds\,\int_{t}^{t +\Delta t} ds'\eta(s) \eta(s') + \cdots \nonumber
\end{eqnarray}
Therefore, the average $\Delta x(t)$ over the noise is
\bb \langle
\Delta x(t) \rangle = \alpha g g' \int_{t}^{t +\Delta t}
ds\,\int_{t}^{t +\Delta t} ds'\, \delta(s-s').
\label{deltax}
\en
The integral
\begin{eqnarray} \int_{-\infty}^{t} ds'
\, \delta(s-s') &=& \left \{ \begin{array}{cc}
                                            0 & s > t \\
                                            1 & s < t
                                           \end{array} \right. \\
                           &=& \theta(t-s),
\end{eqnarray}
defines the Heaviside unit step function, $\theta(t)$, and Eq.\
(\ref{deltax}) becomes
\begin{eqnarray}
\langle \Delta x(t) \rangle
&=& \alpha g g' \int_{t}^{t +\Delta t} ds\,
\left [ \theta( t +\Delta t - s) - \theta(t- s) \right ] \nonumber \\
&=& \alpha\,g g' \Delta t.
\label{method1}
\end{eqnarray}
On the other hand, we can directly integrate Eq.\ (\ref{geta}) to
obtain\bb \Delta x(t) =  \int_{t}^{t +\Delta t}
ds\,g[x(s)]\,\eta(s). \en Expanding $g[x(s)]$ as\bb g[x(s)] =
g[x(t) + \Delta x(s)] = g[x(t)] + g'[x(t)] \Delta x(s)  + \cdots,
\en we find\bb \Delta x(t) = g \int_{t}^{t +\Delta t} \eta(s) + g
g' \int_{t}^{t +\Delta t} ds\,\int_{t}^{s} ds' \eta(s) \eta(s').
\en Note that the upper limit of integration for $s'$ is different
from that in Eq.\ (\ref{deltax}). Therefore, we find \bb \langle
\Delta x(t) \rangle = \theta(0) \, g g' \Delta t. \en Comparing
this with Eq.\ (\ref{method1}), we conclude that $\theta(0)
=\alpha$.

\section{Equivalence between Eq.\ (\ref{path1}) and Eq.\ (\ref{conditional})}
\label{app:equi}

In this Appendix, we demonstrate that the conditional probability
given in Eq.\ (\ref{path1}) and its exponentiated form, Eq.\
(\ref{conditional}), are indeed equivalent. Note that the latter
expression involves a subtle step which is required for the
construction of the path integral in Sec.\ \ref{sec:pathintegral}.
Therefore, it is crucial to confirm Eq.\ (\ref{conditional}) is
correct at least to order of $\Delta t$.  We have already checked
that the correct Fokker-Planck Equation, Eq.\ (\ref{fpe1}), can be
derived from Eq.\ (\ref{conditional}) in Sec.\
\ref{sub:derivation}. Here, we check that the normalization
condition:\bb \int dx_{i} P(x_i\,t_i|x_{i-1}\,t_{i-1}) =1,
\label{norm}
\en for general $\alpha$ is satisfied by Eq.\ (\ref{conditional}).
First, let us check Eq.\ (\ref{norm}) is true for Eq.\
(\ref{path1}). For simplicity, we set $f(x)=0$.  We have
\begin{widetext}\begin{eqnarray} \int dx_i P(x_i t_i |x_{i-1}
t_{i-1})
 = \int { d\Delta x_i \over \sqrt{2 \pi \Delta t } g_i}
\,e^{- { \Delta t  \over 2 g_i^2 }\left ( {\Delta x_i \over \Delta
t} \right )^2}
 \left [ 1 - \alpha\, { g_i' \over g_i }\,\Delta x_i \right ]
 = \int { d w \over \sqrt{2 \pi} g_i}
\left [ 1 - \alpha\, { g_i' \over g_i } \sqrt{\Delta t}\,w \right]
e^{- { w^2  \over 2 g_i^2 }}
\label{a2}
\end{eqnarray}
where we have made a change of variable: $\Delta x_i =
\sqrt{\Delta t}\,w$.  Remembering that $g_i = g[x_{i-1} + \alpha
\sqrt{\Delta t}\,w ]$ and expanding them in Eq.\ (\ref{a2}) in
power of $\Delta t$, we have

\begin{eqnarray}
&&\int dx_i P(x_i t_i |x_{i-1} t_{i-1}) \nonumber \\
&&= \int { d w \over \sqrt{2 \pi} g }  e^{- { w^2  \over 2 g^2 }}
\left [ 1 - {2 \alpha g' \over g } \sqrt{\Delta t}\,w + \left ( {
3\alpha^2 g'^2 \over  g^2} - { 3 \alpha^2 g'' \over 2g} \right  )
\Delta t\,w^2 + { \alpha g' \over g^3} \, \sqrt{\Delta t}\,w^3 +
\left ( { \alpha^2 g'' \over 2 g^3}  - { 7 \alpha^2 g'^2 \over  2 g^4}\right )\, w^4 \right.  \nonumber \\
&& \left. \phantom{\left ({ 7\alpha^2 g'^2 \over  2 g^2} \right
)}
+ { \alpha^2 g'^2 \over 2 g^6 }\,\Delta t\,w^6 + \cdots  \right ] \label{expansion} \\
&& = 1+ {\cal O}[\Delta t^{3/2}],
\end{eqnarray}
where $g \equiv g(x_{i-1})$.  Thus, Eq.\ (\ref{path1}) indeed
satisfies the normalization condition, which is hardly surprising
since it must be true by construction. Now, let us check the
exponentiated from, Eq.\ (\ref{conditional}). We have
\begin{eqnarray}
&& \int dx_i P(x_i t_i |x_{i-1} t_{i-1}) = \int { d\Delta x_i
\over \sqrt{2 \pi \Delta t } g_i} \,e^{- { \Delta t  \over 2 g_i^2
}\left ( {\Delta x_i \over \Delta t} + \alpha g_i g_i'\right )^2}
= \int { d w \over \sqrt{2 \pi} g_i} e^{- { \Delta t  \over 2
g_i^2 }
\left ( {w \over \sqrt{\Delta t}} + \alpha g_i g_i'\right )^2}, \nonumber \\
&& = \int { d w \over \sqrt{2 \pi} g }  e^{- { w^2  \over 2 g^2 }}
\left [ 1 - {1 \over 2}  \alpha^2 g'^2  - {2 \alpha g' \over g }
\sqrt{\Delta t}\,w  + \left ( { 7\alpha^2 g'^2 \over  2 g^2}- { 3
\alpha^2 g'' \over 2g} \right  ) \Delta t\, w^2 + { \alpha g'
\over g^3} \, \sqrt{\Delta t}\, w^3 +
\left ( { \alpha^2 g'' \over 2 g^3}  - { 7 \alpha^2 g'^2 \over  2 g^4}\right )\, w^4 \right.  \nonumber \\
&& \left. \phantom{\left ({ 7\alpha^2 g'^2 \over  2 g^2} \right
)}
+ { \alpha^2 g'^2 \over 2 g^6 }\,\Delta t\,w^6  + \cdots \right ] \label{expansion2}\\
&& = 1+ {\cal O}[\Delta t^{3/2}].
\end{eqnarray}
\end{widetext}
Thus, Eq.\ (\ref{conditional}) also satisfies the normalization
condition. We note in passing that although the expansions, Eqs.\
(\ref{expansion}) and (\ref{expansion2}) are different, they both
give one at the end result to the lowest order and the next order
term is of the order $\Delta t^{3/2}$.

\section{alpha-dependent perturbation theory for the model system with $D(x)= D_0 ( 1+ \gamma x^2) $}
\label{app:perturbation}

\begin{figure}
    \resizebox{2.5in}{!}{\includegraphics{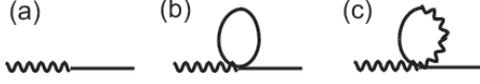}}
    \caption{\label{fig:diagrams2} Diagrams that contribute to the self-energy $\Sigma(\omega)$.}
\end{figure}

In this Appendix, we carry out a first order perturbation
calculation for general $\alpha$ of the model system studied in
Sec.\ \ref{subsec:perturbation} in which $D(x)= D_0 ( 1+ \gamma
x^2) $ in order to clarify some subtle issues associated with the
$\alpha$-convention. It is straightforward to work out the action
in Eq.\ (\ref{path2}). Up to an irrelevant constant, we
have\begin{eqnarray}
{\cal S}_0 &=& \int dt \left [ -D_0 \xh^2 + \xh \left ( \partial_t x + \Gamma_0 k x \right ) \right ] \nonumber \\
{\cal S}_I &=&  \Gamma_0 \gamma \int dt \left [ 2 (2 \alpha -1) x
\xh + { k \over T}\,x^3 \xh
- \,x^2 \xh^2 \right. \nonumber \\
&-& \left. 3\,\alpha\,{ k \over T}\, x^2 \right ]. \nonumber
\end{eqnarray}
Note that there are more diagrams to evaluate than there are for
$\alpha = 0$. Consider first the propagator $G(t,t')= \langle x(t)
\xh(t')\rangle $, which can be written as $G^{-1}(\omega) =
G_0^{-1}(\omega) + \Sigma(\omega)$, where the diagrams for the
self-energy are displayed in Fig.\ \ref{fig:diagrams2}.  Note that
the closed loop diagram c in Fig.\ \ref{fig:diagrams2} contains
$G_0(t=0)$ which must be set to $\alpha$ as explained in Sec.\
\ref{subsec:correlation}. We find
\begin{eqnarray}
\Sigma(\omega) &=& 2(2 \alpha -1) D_0 \gamma  + 3 D_0 \gamma  - 4 D_0 \gamma G_0(0) \nonumber \\
&=& D_0 \gamma, \nonumber
\end{eqnarray}
which agrees with the calculation for $\alpha =0$. Note that the final
result is independent of $\alpha$ as it should be. To first order,
the noise $D(\omega)$ renormalizes exactly the same way as in the
$\alpha =0$ calculation. However, the physical response function
is different.  It is given by Eq.\ (\ref{physresponse}) with an
extra $\alpha$ dependent term:\begin{eqnarray} \chi_{xx}(t, t')
&=&  \Gamma_0 \left \langle\,x(t)\,\xh(t')
 \left [\, 1+ \gamma\,x^2(t')\,\right ] e^{-{\cal S}_I} \right \rangle_0  \nonumber \\
 &-& {\alpha \over T} \left \langle\,x(t) [ 2\,\gamma D_0 x(t')] e^{-{\cal S}_I} \right \rangle_0 \nonumber \\
 &=& \Gamma_0 \left ( 1+ {\gamma T \over k} \right ) G(t,t') + 2 \alpha \Gamma_0  \gamma C_0(t,t') \nonumber \\
 &-& 2 \alpha \Gamma_0  \gamma C_0(t,t') + {\cal O}(\gamma^2)  \nonumber \\.
 &=& \Gamma_0 \left (1+ {\gamma T \over k} \right ) G(t,t'). \nonumber
\end{eqnarray}
Without the cancellation of the $\alpha$ dependent terms,
$\chi_{xx}(t, t')$ would not have been causal.

\begin{thebibliography}{0}

\bibitem{frey}
For a recent review, see E.\ Frey and K.\ Kroy, Ann.\
Phys.(Leipzig) {\bf 14}, 20 (2005).

\bibitem{vankampen1}
N.G. van Kampen, {\em Stochastic Processes in Physics and
Chemistry.} (North-Holland, Amsterdam 1992).

\bibitem{risken}
H.\ Risken, {\em The Fokker-Planck Equation.} (Springer-Verlag,
NY, 1989).

\bibitem{gardiner}
C.W.\ Gardiner, {\em Handbook of Stochastic Methods for Physics,
Chemistry and the Natural Sciences.} (Springer-Verlag, NY, 1983).

\bibitem{justin}
J.\ Zinn-Justin, {\em Quantum Field Theory and Critical
Phenonmena}, 4th ed.\ (Oxford University Press, NY, 2003).

\bibitem{cugliandolo}
L.F.\ Cugliandolo, in {\em Slow relaxations and nonequilibrium
dynamics in condensed matter}, (Springer, NY, 2003).

\bibitem{biology}
J.D.\ Murray, {\em Mathematical Biology}, 3rd ed.\ (Springer
Verlag, Heidelberg, 2002).

\bibitem{lubensky}
P.M. Chaikin and T.C. Lubensky, {\em Principles of Condensed
Matter Physics} (Cambridge University Press, New York, 1995).

\bibitem{cell}
J.\ Howard, {\em Mechanics of Motor Proteins and the
Cytoskeleton}, (Sinauer Associates, 2001).

\bibitem{morse}
D.C.\ Morse, Advances in Chem.\ Phys.\, {\bf 128}, 65 (2004).

\bibitem{colloidfrench}
P.\ Lancon, G.\ Batrouni, L.\ Lobry, and N.\ Ostrowsky, Europhys.\
Lett.\ {\bf 54}, 28 (2001); Physica A {\bf 304}, 65 (2002).

\bibitem{dna}
J.C.\ Neto, R.\ Dickman, O.N.\ Mesquita, Physica A {\bf
345}, 173 (2005); E.\ Goshen, W.Z.\ Zhao, G.\ Carmon, S.\ Rosen,
R.\ Granek, M.\ Feingold, Phys.\ Rev.\ E, {\bf 71}, 061920
(2005).

\bibitem{seifert}
V.\ Blickle, T.\ Speck, L.\ Helden, U.\ Seifert, and C.\ Bechinger, Phys.\ Rev.\ Lett.\ {\bf 96},
070603 (2006).

\bibitem{quake}
J.-C.\ Meiners and S.R.\ Quake, Phys.\ Rev.\ Lett.\ {\bf 82}, 2211
(1999); S.\ Henderson, S.\ Mitchell, and P.\ Bartlett, Phys.\
Rev.\ E {\bf 64}, 061403 (2001).

\bibitem{bruinsma}
T.\ Bickel and R.\ Bruinsma, Biophys.\ J.\ {\bf 83}, 3079 (2002).

\bibitem{cai}
W.\ Cai and T.C.\ Lubensky, Phys.\ Rev. E {\bf 52}, 4251 (1995).

\bibitem{crocker}
J.C.\ Crocker and D.G.\ Grier, Phys.\ Rev.\ Lett.\ {\bf 73}, 352
(1994); J.\ Colloid Interface Sci.\ {\bf 179}, 298 (1996).

\bibitem{ermak}
D.L.\ Ermak and J.A.\ McCammon, J.\ Chem.\ Phys.\ {\bf 69}, 15
(1978).

\bibitem{sancho}
J.M.\ Sancho, M.\ San Miguel, and D.\ D{\" u}rr, J.\ Stat.\ Phys.\ {\bf
28}, 291 (1982).

\bibitem{doi}
M.\ Doi and S.F.\ Edwards, {\em The Theory of Polymer Dynamics}
(Clarendon Press, Oxford, 1994).

\bibitem{vankampen2}
N.G. van Kampen, J.\ Stat.\ Phys.\ {\bf 24}, 175 (1981).

\bibitem{arnold}
P.\ Arnold, Phys.\ Rev.\ E {\bf 61}, 6091 (2000); Phys.\ Rev.\ E
{\bf 61}, 6099 (2000).

\bibitem{oksendal}
B.\ {\O}ksendal, {\em Stochastic Differential Equations.} (Springer,
NY, 2000).

\bibitem{mark}
M.J.\ Schnitzer, Phys.\ Rev.\ E {\bf 48}, 2553 (1993).

\bibitem{crocker2}
J.C.\ Crocker, J.\ Chem.\ Phys.\ {\bf 106}, 2837 (1997).

\bibitem{verma}
R.\ Verma, J.C.\ Crocker, T.C.\ Lubensky, and A.G.\ Yodh,
Macromolecules {\bf 33}, 177 (2000).

\bibitem{dogic}
M.P.\ Lettinga, E.\ Barry, and Z.\ Dogic, Europhys.\ Lett.\ {\bf
71}, 692 (2005).

\bibitem{brenner}
H.\ Brenner, Chem.\ Eng.\ Sci.\ {\bf 16}, 242 (1961).

\bibitem{numerics}
For all numerical simulations done in this paper, we employ the
Miltstein scheme:\begin{eqnarray}
x_{n+1} &=& x_n + f(x_n) h + g(x_n) \Delta W_n \nonumber \\
&+& {1 \over 2}\,g(x_n)g'(x_n)\left [ (\Delta W_n)^2 - h \right ],
\nonumber
\end{eqnarray}
where $h$ is a time step, and $\Delta W_n$ is a Gaussian
distributed random variable with zero mean and variance of
$\sigma^2 = h$. See, for example, P.E.\ Kloeden and E. Platen,
{\em Numerical Solution of Stochastic Differential Equations.}
(Springer, NY, 1995).

\bibitem{libchaber}
L.P.\ Faucheux and A.J.\ Libchaber, Phys.\ Rev.\ E {\bf 49}, 5158
(1994).

\bibitem{grier}
E.R.\ Dufresne, D.\ Altman, and D.G.\ Grier, Europhys.\ Lett.\ {\bf
53}, 264 (2001).

\bibitem{MSR}
P.C.\ Martin, E.D.\ Siggia, H.A.\ Rose, Phys.\ Rev.\ A {\bf 8},
423 (1973).

\bibitem{phythian}
R.\ Phythian, J.\ Phys.\ A: Math.\ Gen.\ {\bf 10}, 777 (1977); B.\
Jouvet and R.\ Phythian, Phys.\ Rev.\ A {\bf 19}, 1350 (1979).

\bibitem{graham}
R.\ Graham, in {\em Springer Tracts in Modern Physics in Solid
State}, Vol. 66 (Springer-Verlag, NY, 1973).

\bibitem{janssen}
H.K.\ Janssen, Z.\ Phys.\ B {\bf 23}, 377 (1976).

\bibitem{reichman}
K.\ Miyazaki and D.R.\ Reichman, J.\ Phys.\ A: Math.\ Gen.\ {\bf
38}, L343-L355 (2005).


\end{thebibliography}
\end{document}